\documentclass[fleqn,usenatbib,useAMS]{mnras}
\usepackage[T1]{fontenc}
\usepackage{ae,aecompl}
\usepackage{graphicx}
\usepackage{amsmath}
\usepackage{amssymb}

\title[{\it RXTE} observations of EXO~2030+375]
{Decade long {\it RXTE} monitoring observations of Be/X-ray binary pulsar EXO~2030+375}

\author[Epili et al.]
{Prahlad Epili\thanks{E-mail: prahlad@prl.res.in}, Sachindra Naik\thanks{snaik@prl.res.in},
Gaurava K. Jaisawal\thanks{gaurava@prl.res.in} and Shivangi Gupta\thanks{shivangi@prl.res.in} \\
Astronomy and Astrophysics Division, Physical Research Laboratory, Navrangapura, 
Ahmedabad - 380009, Gujarat, India}

\date{}

\pubyear{2017}

\begin{document}
\label{firstpage}
\pagerange{\pageref{firstpage}--\pageref{lastpage}}
\maketitle

\begin{abstract}

We present a comprehensive timing and spectral studies of Be/X-ray binary pulsar EXO~2030+375
using extensive {\it Rossi X-ray Timing Explorer} observations from 1995 till 2011, covering 
numerous Type~I and 2006 Type~II outbursts. Pulse profiles of the pulsar were found to be 
strongly luminosity dependent. At low luminosity, the pulse profile consisted of a main peak
and a minor peak that evolved into a broad structure at high luminosity with a significant
phase shift. A narrow and sharp absorption dip, also dependent on energy and luminosity, 
was detected in the pulse profile. Comparison of pulse profiles showed that the features at a 
particular luminosity are independent of type of X-ray outbursts. This indicates that the 
emission geometry is solely a function of mass accretion rate. The broadband energy 
spectrum was described with a partial covering high energy cutoff model as well as a physical 
model based on thermal and bulk Comptonization in accretion column. We did not find any 
signature of cyclotron resonance scattering feature in the spectra obtained from all the 
observations. A detailed analysis of spectral parameters showed that, depending on source 
luminosity, the power-law photon index was distributed in three distinct regions. 
It suggests the phases of spectral transition from sub-critical to super-critical regimes 
in the pulsar as proposed theoretically. A region with constant photon index was also 
observed in $\sim$(2-4)$\times$10$^{37}$~erg~s$^{-1}$ range, indicating critical luminosity 
regime in EXO~2030+375. 

\end{abstract}

\begin{keywords}
stars: neutron -- pulsars: individual: EXO~2030+375 -- X-rays: stars.
\end{keywords}

\section{Introduction}

Most of the accretion powered binary X-ray pulsars are among the brightest sources in our Galaxy.
 They belong to the class of high mass X-ray binaries in which a neutron star accretes matter from
 a massive $>$10~M$_\odot$ main-sequence companion. Depending on the evolutionary state of the donor,
 mass transfer from the companion star to the compact object takes place through capture of stellar
 wind or accretion from a huge circumstellar disk around the companion star \citep{Paul2011}.
 Among confirmed high mass X-ray binaries (HMXBs), Be/X-ray binaries (BeXBs) represent about 
two-third population with neutron star as the compact object. The optical companion in these 
systems is a non-supergiant B or O type star that shows emission lines in its spectrum \citep{Reig2011}.
 An excess emission in the infrared band is also observed from these companion stars in BeXBs.
 It is believed that due to rapid rotation, the Be star expels material equatorially forming
 a huge disk, called as circumstellar disk, around it. The observed emission lines and infrared
 excess in the spectrum are attributed to the presence of circumstellar disk around the central Be star.

A neutron star in BeXBs revolves in a wide and moderate eccentric orbit. While passing close to the 
periastron, an abrupt mass accretion from the circumstellar envelope onto the neutron star gives
 rise to strong X-ray outbursts. The intensity during such outbursts increases up to an order of
 magnitude than the quiescent phase. BeXBs generally show periodic or normal (Type~I) X-ray outbursts
 that occur at the periastron passage of the neutron star. These outbursts cover a small fraction
 of the orbit ($<$20--30\%) and last for a few days to weeks \citep{Stella1986}. Another class of
 X-ray outbursts such as giant outbursts (Type~II) are also seen from the neutron stars in BeXBs.
 These outbursts cover a significant fraction or multiple orbits lasting for several weeks to months.
 Type~II X-ray outbursts are quite rare and independent of the orbital phase or periastron passage
 of the binary. During normal and giant X-ray outbursts, the luminosity of the pulsar generally
 reaches up to $\le$10$^{37}$ and 10$^{38}$~erg s$^{-1}$, respectively \citep{Okazaki2001}. 

The spin period of BeXB pulsars ranges from a few seconds to about thousand seconds. During outbursts,
 change in spin period of the pulsars has also been observed. This occurs because of torque transfer
 from accreting material to the neutron star. Accretion powered pulsars usually show broadband emission
 ranging from soft to hard X-rays. It is interpreted as due to thermal or bulk Comptonization of
 seed photons from the hot spots on the neutron star surface across the accretion column \citep{Becker2007}.
 Apart from the continuum, the energy spectrum of pulsars also shows the presence of several other
 components such as soft X-ray excess, iron fluorescence emissions, cyclotron resonance scattering
 features (CRSF) etc. Detection of CRSFs in the pulsar spectrum provides a direct and most accurate
 method for the estimation of surface magnetic field of neutron stars. These features are originated
 due to the resonant scattering of photons with electrons in quantized energy levels in the presence
 of strong magnetic field. Spectrum and pulse profile (representation of beam function) of the pulsars
 can change depending on the mass accretion rate, accretion geometry and physical processes occurring 
close to the neutron star. For a detailed description on the properties of transient HMXB pulsars, 
refer to articles by \citet{Paul2011} and \citet{Caballero2012}.                
 
Be/X-ray binary pulsar EXO~2030+375 was discovered during a giant X-ray outburst in 1985, with {\it EXOSAT}
 observatory \citep{Parmar1989b}. The observations during the same outburst revealed the pulsating nature of
 the neutron star with a spin period of 42~s along with an orbital modulation of 44.3-48.6 days. The transient
 nature of the pulsar was revealed during these observations. The 1-20 keV luminosity of the pulsar was observed
 to change by a factor of $\ge$2500 from quiescence over a duration of 100 days. A significant spin-up 
(${-P}/{\dot{P}}\sim30$~yr) of the pulsar was also observed during the {\it EXOSAT} observations \citep{Parmar1989b},
 suggesting the presence of an accretion disk. Using optical and infrared observations, the counterpart of
 compact object was discovered as a highly reddened B0 Ve star located at a distance of 7.1 kpc 
\citep{Motch1987,Coe1988, Wilson2002}. \citet{Stollberg1997} derived orbital parameters of the binary
 system by using long term {\it BATSE} monitoring data. The orbital period was determined more precisely
 and found to be 46 days. Strong luminosity dependent pulse profile was seen during the 1985 outburst 
\citep{Parmar1989a}. At higher luminosity, the pulse profile of the pulsar was characterized by two peaks
 (main peak and a minor peak) separated by a phase difference of $\sim$0.5. The strength of these two 
peaks was found to alter when the source luminosity decreased by a factor of $\sim$100. This observed
 change in strength and structure of each of the peaks in the pulse profile with luminosity was attributed
 to the change in the emission beam pattern e.g. from fan beam to pencil beam geometry \citep{Parmar1989b}.

EXO~2030+375 is a unique BeXB pulsar which shows regular Type~I X-ray outbursts at the periastron
 passage \citep{Wilson2002}. At the peak of the Type~I outbursts observed with the {\it RXTE} in
 1996--2006, source flux was approximately 100~mCrab. A correlation between the spin frequency 
and luminosity indicated that the pulsar was spinning-up during brighter outbursts in 1992-1994. 
On the other hand, a spin-down trend was observed during low luminous outbursts in 1994--2002 
\citep{Wilson2002, Wilson2005}. EXO~2030+375 was caught into a giant outburst in June 2006 with 
source flux peaking up to $\sim$750 mCrab \citep{Krimm2006}. During this giant outburst which 
lasted for about 140 days, the neutron star showed a remarkable spin-up behaviour \citep{Wilson2008}.
 After this outburst, many intense Type~I outbursts ($\le$300 mCrab) were detected for a number of
 orbits till the source settled to its regular mode. Since early 2015, however, the pulsar had 
undergone to a period of low activity for more than a year. The peak flux drastically went down
 during this phase and hardly any X-ray enhancement was seen at the expected periastron epochs 
\citep{Fuerst2016}. Recent observations with {\it Swift}/XRT and {\it NuSTAR} after March 2016,
 however, confirmed the recurrence of X-ray activity (Type~I X-ray outbursts) along with spin-down
 trend in the neutron star \citep{Kretschmar2016}. 

The energy spectrum of pulsar obtained from the 1985 giant outburst was described by a power-law model 
along with thermal blackbody component at 1.1 keV (\citealt{Sun1994} and reference therein). However, 
an absorbed power-law modified with high energy cutoff model was widely used in later observations of
 EXO~2030+375 during Type~I and Type~II outbursts \citep{Reig1999, Wilson2008}. Apart from the 6.4~keV
 iron florescence emission line, detection of cyclotron absorption line was reported at three different
 energies such as $\sim$11, 36 and 63 keV in pulsar spectra obtained from {\it  RXTE} and {\it INTEGRAL}
 observations during different X-ray outbursts \citep{Reig1999, Wilson2008, Klochkov2008}. However, 
{\it Suzaku} observations during 2007 May-June and 2012 May Type~I outbursts did not confirm the presence
 of any such features in the pulsar spectrum \citep{Naik2013, Naik2015}. Above {\it Suzaku} observations
 also showed some other interesting aspects. 
Along with iron lines, several emission lines were also detected in the spectrum. It was the first time 
when an absorption dip was detected in the pulse profile up to as high as $\sim$70~keV \citep{Naik2013}.
 This was explained as due to the presence of additional dense matter (partial absorber) at certain phases
 of the pulsar. A peculiar narrow absorption dip was also detected in soft X-ray pulse profile obtained
 from {\it XMM-Newton} observation in May 2014 at a luminosity of $\sim$$10^{36}$~erg~s$^{-1}$ \citep{Ferrigno2016}.
 This feature was interpreted as the effect of self absorption from accretion mount onto the neutron star surface.

\begin{table*}
\centering
\caption{Log of {\it RXTE}/PCA observations of the pulsar EXO~2030+375 during 
Type~I and Type~II outbursts.}
\begin{tabular}{ccccc}
\hline\\
 Year of Observations    &Proposal           &No. of Obs.  &Time range &On Source time \\

  	 	         &ID		     &(IDs)        &(MJD)	    &(ksec)\\
\hline
\\
1996 Jul		&P10163			&18	&50266.55 -- 50274.56	&67.42 \\
1998 Jan		&P30104			&2	&50825.02 -- 50827.77	&37.94 \\
2002 Jun		&P70074			&21	&52431.97 -- 52441.47	&76.91 \\
2003 Sep		&P80071			&15	&52894.44 -- 52898.32	&145.87 \\
2005 Jun -- 2006 Feb	&P91089			&52	&53540.78 -- 53776.28	&145.57 \\
2006 Mar -- 2006 Nov	&P92067, P91089, P92066 &143	&53816.96 -- 54069.97	&342.34 \\
2006 Dec -- 2007 Jun	&P92422			&147	&54070.95 -- 54279.51	&221.91 \\
2007 Jun -- 2008 Oct	&P93098			&79	&54280.56 -- 54749.53	&202.41 \\
2008 Dec -- 2009 Oct    &P94098			&40	&54830.20 -- 55114.49	&92.55 \\
2010 Jan -- 2010 Nov	&P95098			&43	&55197.87 -- 55530.08	&83.79 \\
2011 Jan -- 2011 Nov	&P96098			&46	&55566.06 -- 55895.74	&106.68 \\
\\
\hline
\end{tabular}
\label{log}
\end{table*}

In this paper, we present a detailed study of decade long {\it RXTE} monitoring observations of the pulsar over
a wide range of luminosity. Investigations on the pulse profiles and corresponding spectral parameters were
performed to understand the properties of the pulsar during Type~I and Type~II outbursts.
Along with standard continuum models used to describe the pulsar spectrum, we also used 
a physical model based on thermal \& bulk Comptonization of  infalling plasma in the accretion 
column (BW model; \citealt{Becker2007, Ferrigno2009}) to describe the spectrum of EXO~2030+375. 
We have used this model to understand the physical properties of accretion column across a wide 
range of the pulsar luminosity. Section~2 describes the details of observations and data analysis 
procedures for {\it RXTE} and {\it NuSTAR} observations.  We present the results obtained from timing 
and spectral studies in Section~3. The implication of our results are discussed in Section~4.

\begin{figure*}
\centering
\includegraphics[height=6.4in, width=4.9in, angle=-90]{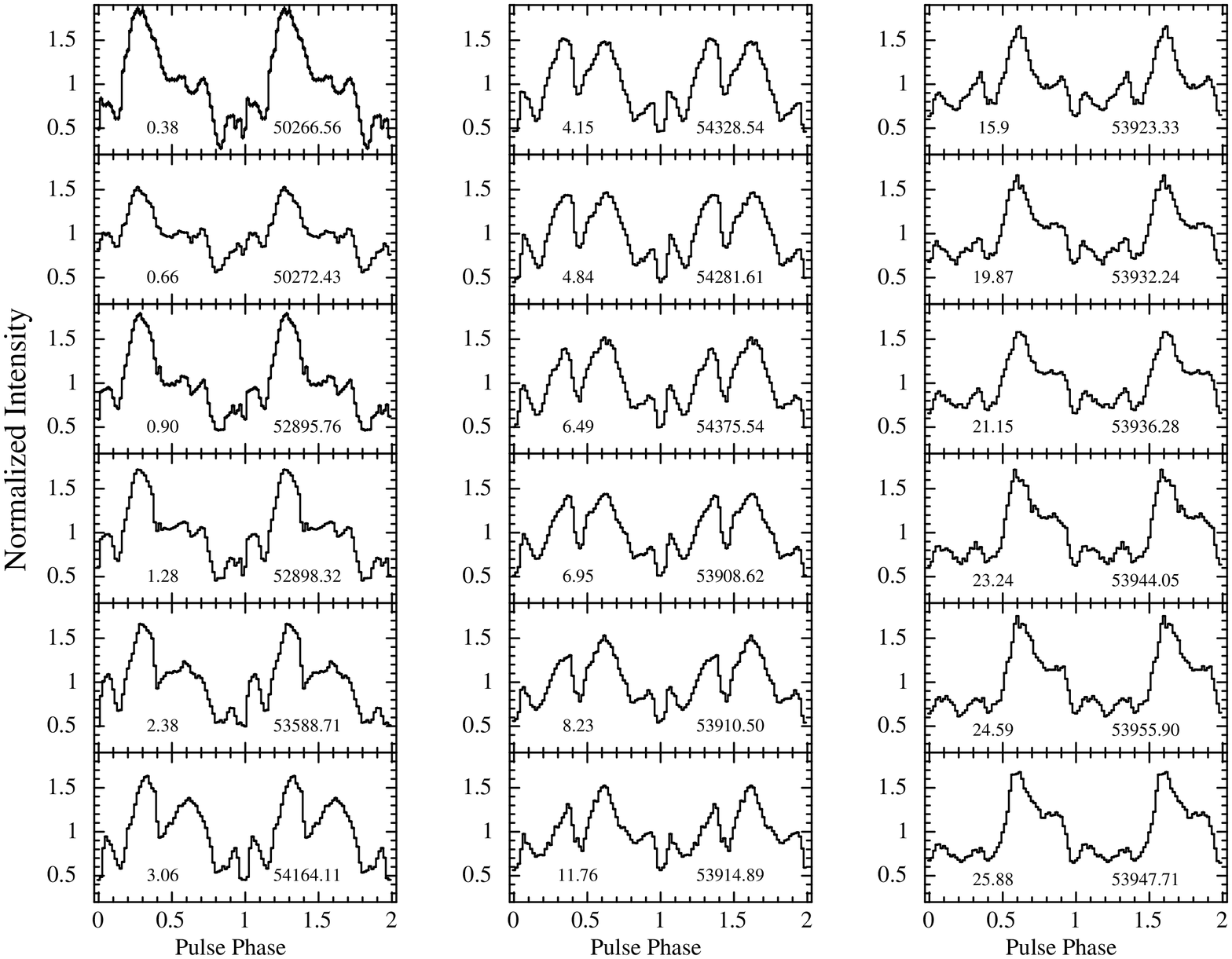}
\caption{Pulse profiles of EXO~2030+375 obtained from Type~I and Type~II outbursts at different luminosities
 in 2-60 keV range. These profiles are generated by folding the {\it RXTE}/PCA light curves at respective
 pulse period estimated separately for each of the observations. The epochs used for folding the light
 curves with corresponding pulse period were considered close to the beginning of the observation and
 adjusted manually to align the pulse profiles for comparison. The numbers quoted in left and right
 side of each panel denote the 3-30~keV luminosity (in units of $10^{37}$~erg~s$^{-1}$) and beginning
 of the corresponding observation (in MJD) of the pulsar, respectively. Two pulses are shown in each
 panel for clarity. The error bars represent 1$\sigma$ uncertainties.}
\label{fig1}
\end{figure*}

\begin{figure*}
\centering
\includegraphics[height=4.in, width=4.3in, angle=-90]{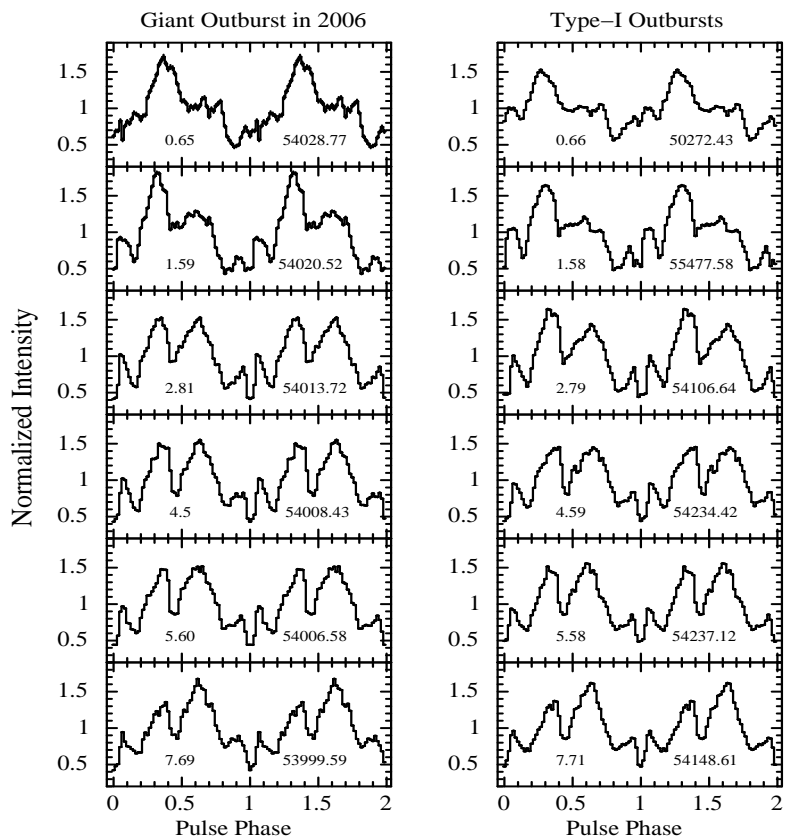}
\caption{Pulse profiles of EXO~2030+375 during the Type~II outburst in 2006 (left panels) and during
 normal Type~I outbursts (right panels) in 2-60 keV range at comparable luminosities. The 3-30 keV
 luminosity of the pulsar (in 10$^{37}$~erg s$^{-1}$ units) and beginning of corresponding
 observation (in MJD) are quoted in left and right side of each panel. Two pulses are shown in each
 panel for clarity. The error bars represent 1$\sigma$ uncertainties.} 
\label{fig2}
\end{figure*}
\begin{figure}
\centering
\includegraphics[height=2.5in, width=3.5in, angle=-90]{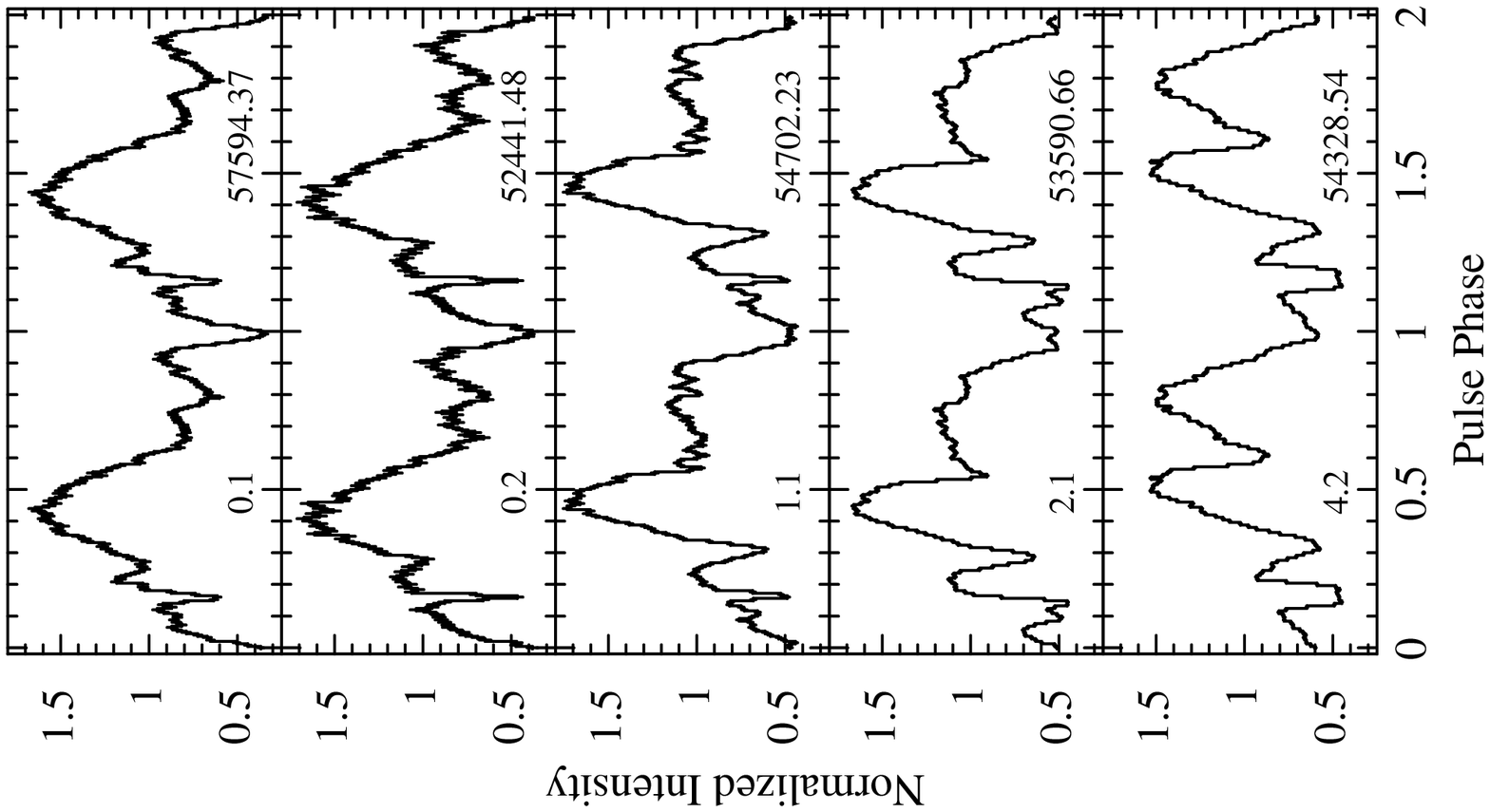}
\caption{A peculiar narrow dip (in 0.1-0.2 phase range) can be seen in the pulse profiles of EXO~2030+375.
 The profile in top panel represents data from the {\it NuSTAR} observation whereas other panels show
 profiles obtained from the {\it RXTE} observations. The 3-30 keV luminosity 
(in unit of 10$^{37}$~erg s$^{-1}$) and the beginning of corresponding observation (in MJD) are quoted
 in left and right sides of each panel, respectively. Two pulses are shown in each panel for clarity.
 The error bars represent 1$\sigma$ uncertainties.} 
\label{fig3}
\end{figure}

\section{Observations and Analysis}
 
EX0~2030+375 has been observed numerous times during {\it RXTE} era i.e. between 1995-2012. These 
observations were performed at multiple epochs during 2006 Type~II and several Type~I outbursts. 
We have analyzed a total of 606 pointing observations for an effective exposure of 
{\it 1.52 million seconds} in the present study to understand timing and spectral properties of 
the pulsar over a span of 15 years. A detail log of relevant observations are given in Table~\ref{log}.
 A {\it Nuclear Spectroscopic Telescope Array} ({\it NuSTAR}) observation of the pulsar performed
 during an extended period of low activity in 2015 is also used in our work.

\subsection{\it RXTE}

Space based X-ray observatory {\it RXTE} was launched on 1995 December 30 in a low earth orbit for
 understanding the physics and dynamics of compact objects. It carried highly timing and moderate
 spectral capability instruments such as Proportional Counter array (PCA; \citealt{Jahoda1996}) 
and High Energy Timing Experiment (HEXTE; \citealt{Rothschild1998}), sensitive in the soft to hard
 X-rays. The all sky monitor (ASM; \citealt{Levine1996}), effective in 1.5-12 keV range was also the
 essential part of the observatory. The PCA detector of {\it RXTE} consisted of an array of five 
identical Proportional Counter Units (PCUs) with a total collecting area of $\sim$6500~cm$^{2}$.
 The individual PCUs had three Xenon layers having two anode chains in each of them. Each PCU was
 sensitive in 2-60 keV range and had a field of view of 1~$\deg$ FWHM \citep{Jahoda2006}. Including
 the Xenon-filled main counter, the detector also had a Propane ``veto'' layer on either sides with 
aluminized Mylar window for background rejection. The hard X-ray unit of {\it RXTE}, HEXTE consisted
 of two main clusters A and B, rocking orthogonally to provide simultaneous measurement of source 
and background. Each cluster was made of four NaI(Tl)/CsI(Na) phoswich scintillation detectors working
 in 15--250 keV range. The total collecting area of both the clusters was $\sim$1600~cm$^{2}$. 

For our analysis, we have utilized publicly available {\it RXTE} observations acquired from 1996 July
 to 2011 November carried out during various Type~I and Type~II X-ray outbursts (see Table~\ref{log}).
 We have mainly used the PCA standard-1 and standard-2 binned mode data for timing and spectral studies
 of the pulsar, respectively. For processing, standard methods are followed by creating appropriate
 good time interval file and filter selection on all available PCUs using HEASoft (version 6.16) package.
 The source light curves were extracted in the 2-60 keV from standard-1 data at 0.125~s time resolution
 using {\it saextrct} task of FTOOLS. Using {\it runpcabackest} command, the corresponding background
 light curves were generated from standard-2  data by using the bright background model provided by
 instrumentation team. Source and background spectra were extracted from standard-2 data by using
 {\it saextrct} task for all these observations. The response matrices were created for each of
 the observations by using {\it pcarsp} task. In addition to PCA, the HEXTE data were also analyzed
 to get hard X-ray spectrum from cluster-B for the observations carried out during 2006 giant outburst
. Standard procedures were followed to extract source and background spectra and corresponding
 response matrices. Dead-time correction was also applied on the HEXTE spectra.

\subsection{\it NuSTAR}

{\it NuSTAR} is the first hard X-ray imaging observatory which was launched in 2012 June \citep{Harrison2013}.
 It consists of two identical grazing angle focusing telescopes FPMA and FPMB, operating in the range of 3-79 keV.
 A target of opportunity observation (ID: 90201029002) was performed for the pulsar on 2015 July 25 for an
 effective exposure of $\sim$57~ks. Though the observation was carried out at an orbital phase where Type~I X-ray
 outburst was expected, there was no significant X-ray activity observed in the {\it Swift}/BAT monitoring light
 curve \citep{Fuerst2016}. Using NUSTARDAS software v1.4.1 of HEASoft, we have reprocessed the data and generated
 barycentric corrected light curves, spectra, and response matrices and effective area files. The source products
 were estimated from a circular region of 120 arcsec around the central source from FPMA and FPMB event data.
 The background light curve and spectra were also accumulated in a similar manner by considering a circular
 region of 90 arcsec away from the source.


\section{Timing analysis}

\subsection {Luminosity dependent pulse profiles}

As described above, source and background light curves in 2-60 keV range were extracted at a time resolution
 of 0.125~s from {\it RXTE}/PCA data. The barycentic correction was applied to the background subtracted 
light curves by using {\it faxbary} task of FTOOLS. The $\chi^2$-maximization technique was used to estimate 
the spin period of the neutron star for all epochs of {\it RXTE} observations used in present analysis. 
We generated pulse profiles of the pulsar by folding the light curves at respective spin periods.
 The epochs used for folding were chosen close to the beginning of the observations and in such a manner
 that all the pulse profiles are aligned at their minima for a better comparison. In order to understand
 the emission geometry, we were interested to explore the  shape of pulse profiles over a wide range of 
source luminosity. The observations obtained from various Type~I outbursts and 2006 giant outburst are
 used in our study. Figure~\ref{fig1} shows the 2-60 keV pulse profiles of the pulsar in increasing trend
 of luminosity, starting from $\sim$3.8$\times$10$^{36}$ to 2.6$\times$10$^{38}$ erg s$^{-1}$. The numbers 
quoted in each panel of Figure~\ref{fig1} represent the 3-30 keV source luminosity (in the unit of 
10$^{37}$ erg s$^{-1}$; first number) and the beginning of the corresponding observation (in MJD; second number).
 The luminosity of the source during each of the observations was calculated based on the spectral fitting
 (see section~\ref{section-spec}). It is clear from the figure that the pulse profiles are strongly dependent 
on source luminosity. At lower luminosity ($\sim$10$^{36}$ erg s$^{-1}$), the pulse profile consisted of a 
significant peak at pulse phase of $\sim$0.3 along with a minor peak at $\sim$0.7 phase. Sharp dip-like features
 were also visible in the profiles below 0.2 phase. With increase in luminosity, the secondary minor peak starts
 evolving into prominence and becomes comparable to the primary peak (at $\sim$0.3 pulse phase) at a luminosity 
of $\sim$(4--7)$\times$10$^{37}$ erg s$^{-1}$. Beyond this, the second peak remains significant whereas the 
strength of first peak gradually decreases and finally disappears from the pulse profile at a luminosity of
 $\ge$1.6$\times$10$^{38}$~erg s$^{-1}$. Multiple absorption dips clearly appeared in the pulse profiles at
 certain pulse phases during the evolution with source luminosity. It is worth mentioning that the shape of
 pulse profiles at extreme ends of observed source luminosity (first and last panels of Figure~\ref{fig1})
 is relatively similar with a significant phase shift.

\subsection {Pulse profiles during Type~I and Type~II outbursts} 

To investigate the changes in observed properties of the pulsar during Type~I and Type~II X-ray outbursts,
 pulse profiles at various epochs were generated and shown in Figure~\ref{fig2}. The data from several 
Type~I outbursts were used at appropriate flux level to compare with the observations at same intensity 
level, carried out during the 2006 June giant outburst. Left panels in the figure show the pulse profiles
 obtained from several observations during the 2006 giant outburst whereas the right panels show the 
profiles obtained from various Type~I outbursts at comparable source luminosity. From the figure, it 
can be noted that (i) shape of pulse profiles are similar at comparable source intensity, irrespective 
of type of X-ray outbursts, (ii) evolution of pulse profiles with luminosity during both type of outbursts
 are also similar as remarked in Figure~\ref{fig1}. This signifies that pulse profiles of EXO~2030+375 
are independent of the type of X-ray outbursts. 

A thorough investigation of pulse profiles of the pulsar obtained from observations performed during
 Type~II outburst was also carried out. The motivation was to probe the evolution and possible change
 in the emission geometry at same intensity during the rising and declining part of the 2006 giant
 outburst. Due to large number of pointings in 2006 June, we were able to trace the pulsar beam
 function at both phases of the outburst. For this, the pulse profiles were generated at luminosity
 ranging from 10$^{37}$ to 10$^{38}$ erg s$^{-1}$. Our results showed that the profiles during the
 rising and declining phases of the outburst are similar at comparable luminosities. This behavior
 was  analogous to as seen in Figure~\ref{fig2}.

\subsection {Peculiar narrow and sharp absorption dip} 

We extracted a light curve with a time resolution of 0.1~s in 3-60 keV range by using {\it NuSTAR}
 observation of the pulsar. As described above, standard data reduction procedure was followed to
 extract the source and background light curves. From the background subtracted and barycentric
 corrected light curve, spin period of the pulsar was estimated by using {\it efsearch} task of
 FTOOLS and found to be 41.2932(2)~s. Using this period, the pulse profile of the pulsar was 
generated by folding the light curve at 128 phase bins and shown in the top panel of Figure~\ref{fig3}.
 Other panels of figure show the profiles from {\it RXTE} observations in the 2-60 keV range 
in increasing order of source luminosity. A narrow and sharp dip like structure in 0.1-0.2 phase 
range can be clearly seen in the profiles when the pulsar luminosity was $\sim$10$^{36}$~erg s$^{-1}$
 ({\it NuSTAR} observation). This peculiar feature was also detected in the pulse profile obtained 
from {\it XMM}-Newton observation of the pulsar during the 2014 May outburst \citep{Ferrigno2016}. 
Figure~\ref{fig3} shows that this peculiar dip is luminosity dependent and can be seen in the profiles
 up to the luminosity $\le$4$\times$10$^{37}$~erg s$^{-1}$. Beyond this luminosity, the feature merges
 into a broader dip in the profile (bottom panel of Figure~\ref{fig3}). The evolution of the peculiar
 feature with energy also showed strong variation that can traced up to 30~keV by using data from 
{\it NuSTAR} observation.


 \begin{figure*}
 \begin{center}$
 \begin{array}{cccc}
 \includegraphics[height=5.5 cm,angle=-90]{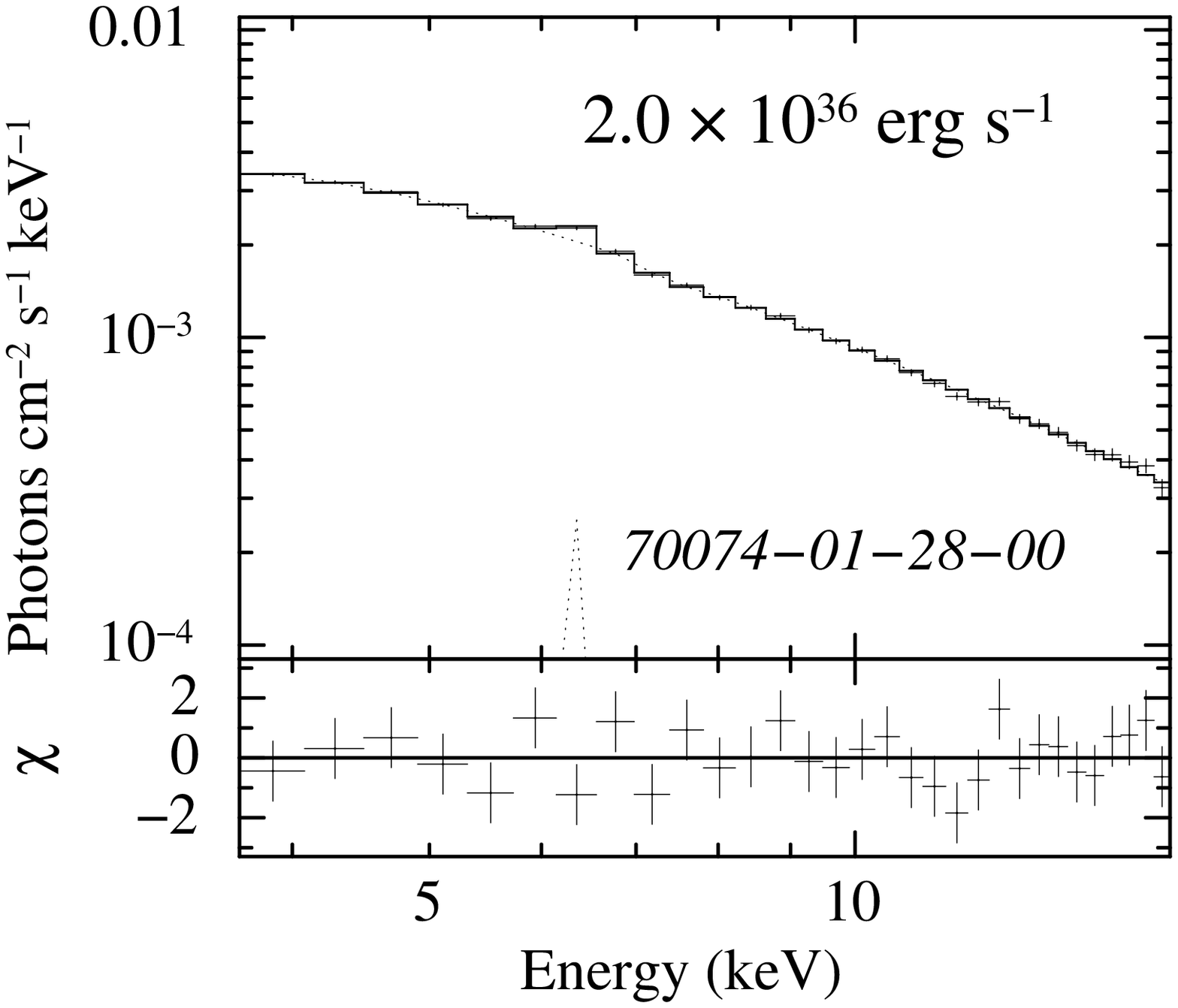} &
 \includegraphics[height=5.5 cm,angle=-90]{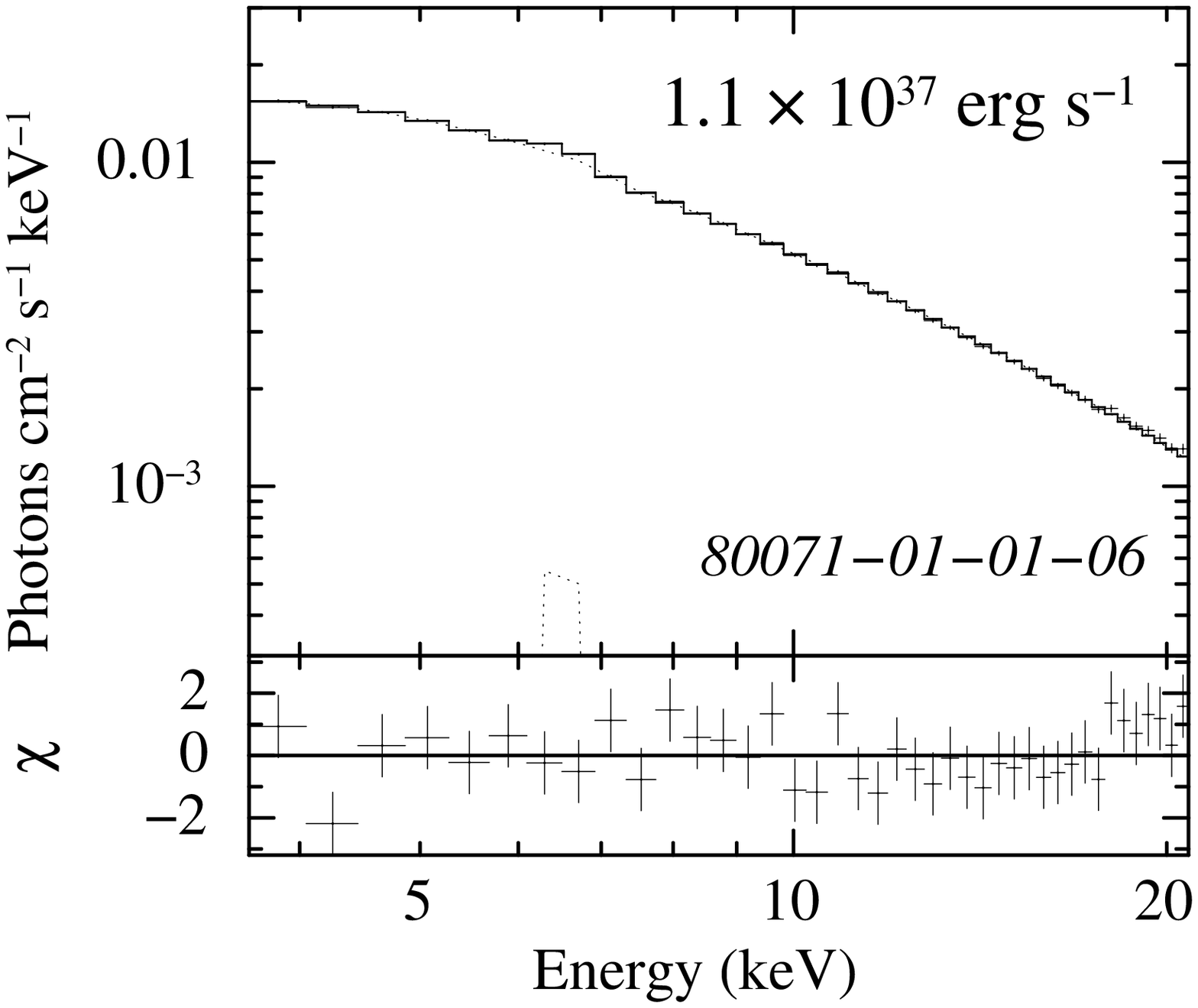} & 
 \includegraphics[height=5.5 cm,angle=-90]{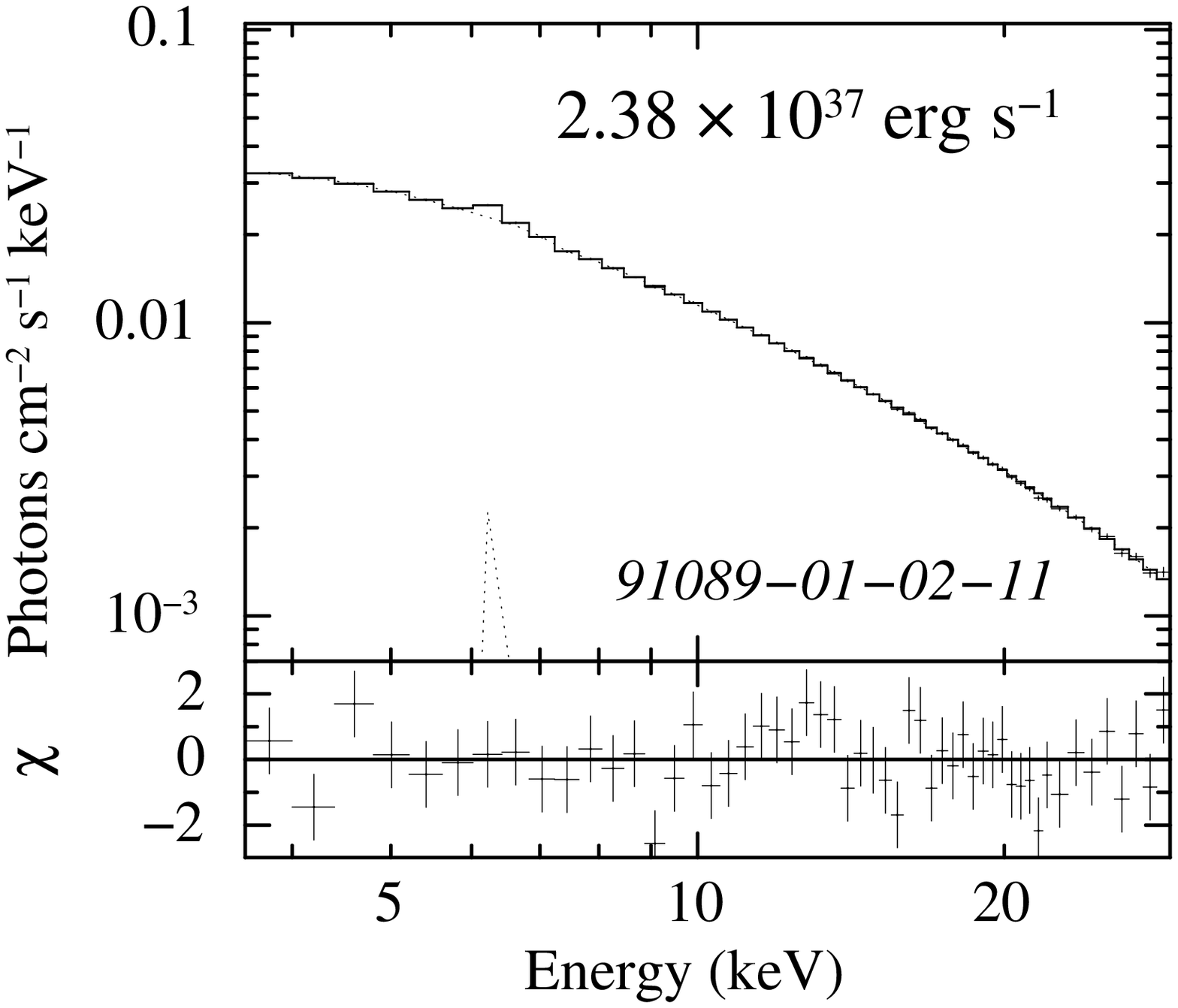} & \\
 \includegraphics[height=5.5 cm,angle=-90]{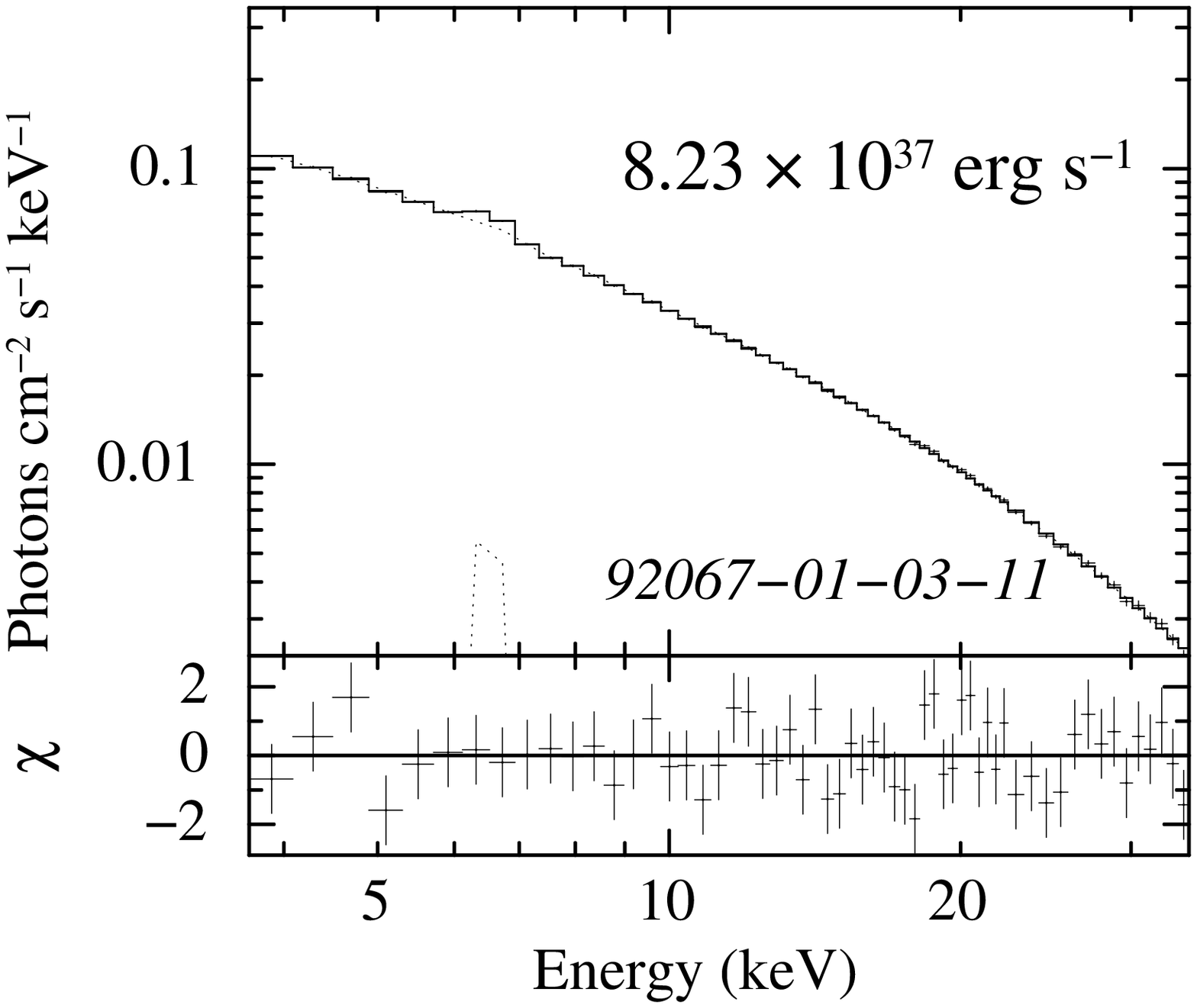} & 
 \includegraphics[height=5.5 cm,angle=-90]{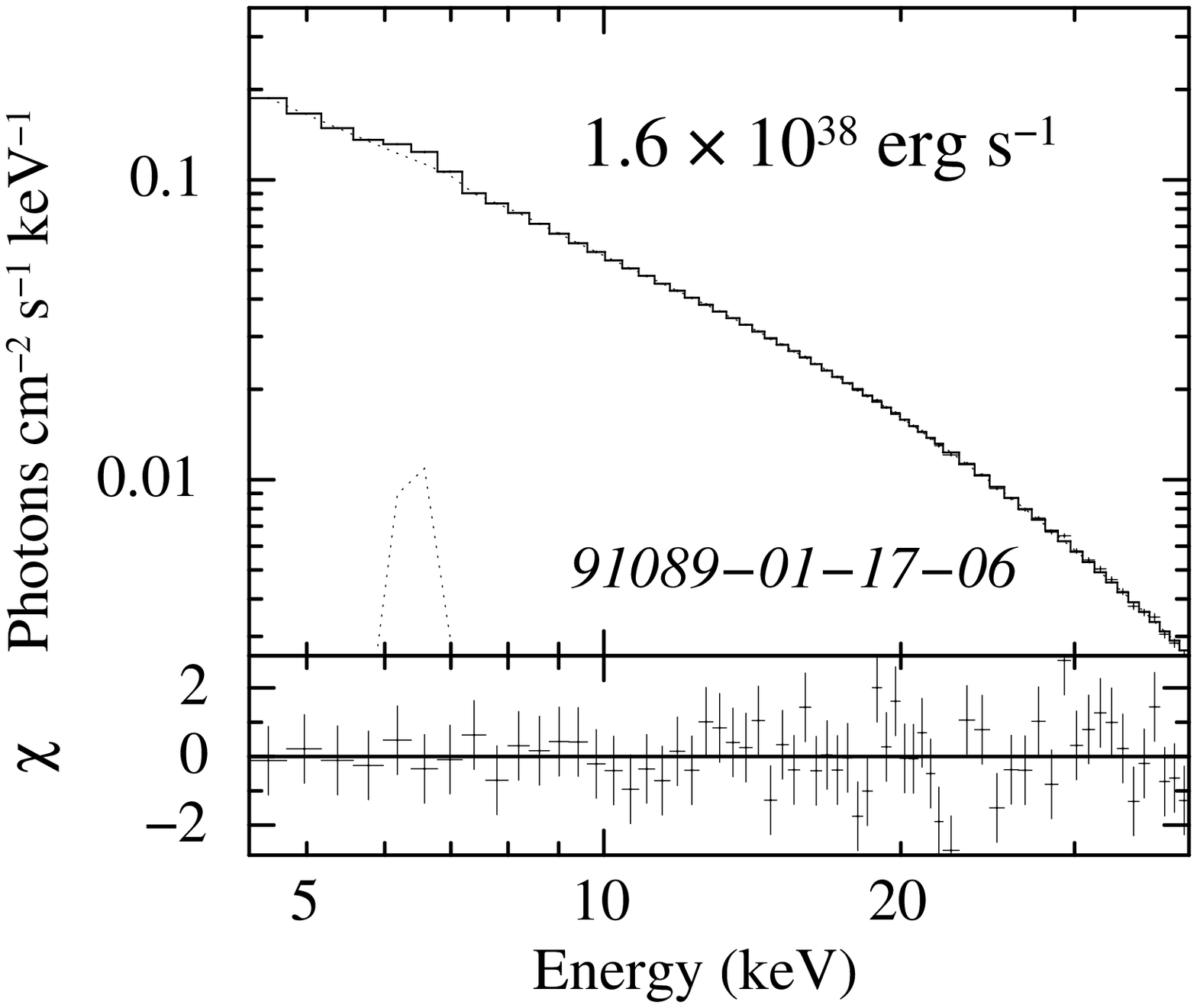} & 
 \includegraphics[height=5.5 cm,angle=-90]{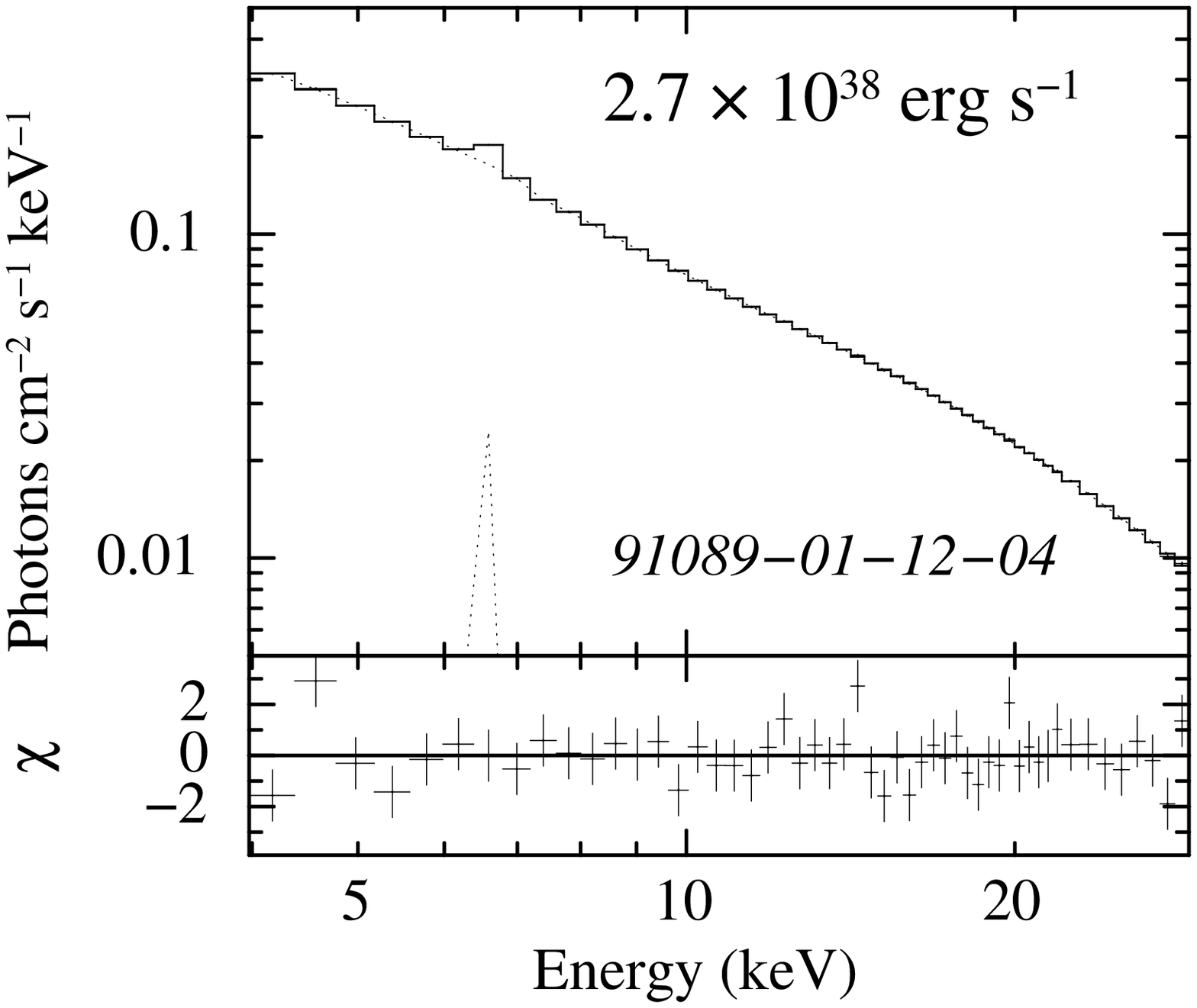} & 
 \end{array}$
 \end{center}
 \caption{Phase-averaged energy spectra of EXO~2030+375 at different luminosity levels, obtained from 
six epochs of {\it RXTE} observations during Type~I and Type~II X-ray outbursts. The spectra were fitted
 with partial covering high energy cutoff model along with an iron emission line at $\sim$6.4 keV.
 The source spectrum and best-fit model are shown in the top panel whereas the contribution of residuals
 to $\chi^2$ at each energy bin are shown in the bottom panel for each epoch of {\it RXTE} observations.
 The observation IDs (in italics) and corresponding source luminosity are quoted in the figure.}
\label{fig4}
\end{figure*}   

\section {Spectral analysis}

\subsection {Phase-averaged spectroscopy}\label{section-spec}

To probe spectral characteristics of the pulsar and corresponding changes with luminosity, we carried 
out spectral studies by using data from {\it RXTE} observations. Source and background spectra were
 extracted by following standard procedure as described in Section~2.1 for all the {\it RXTE} observations.
 Using appropriate background spectra and response matrices, the source spectra from PCA detector were
 fitted by using {\it XSPEC} package. Data in 3--30 keV range were used in the spectral fitting.
 A systematic uncertainty of 0.5\% was added to the data. While fitting the data, we explored several
 continuum models that are used to describe the energy spectrum of accretion powered X-ray pulsars.
 These models are high-energy cutoff power-law, cutoff power-law, negative and positive exponentiation
 cutoff power-law, and with more physical model such as CompTT. In our analysis, we used an absorbed 
power-law model with a high-energy cutoff to describe the continuum spectrum of the pulsar. This model
 has been frequently used to express the broadband spectrum of EXO~2030+375 \citep{Reig1999, Wilson2008}.

\begin{figure}
\centering
\includegraphics[height=3.3in, width=4.2in, angle=-90]{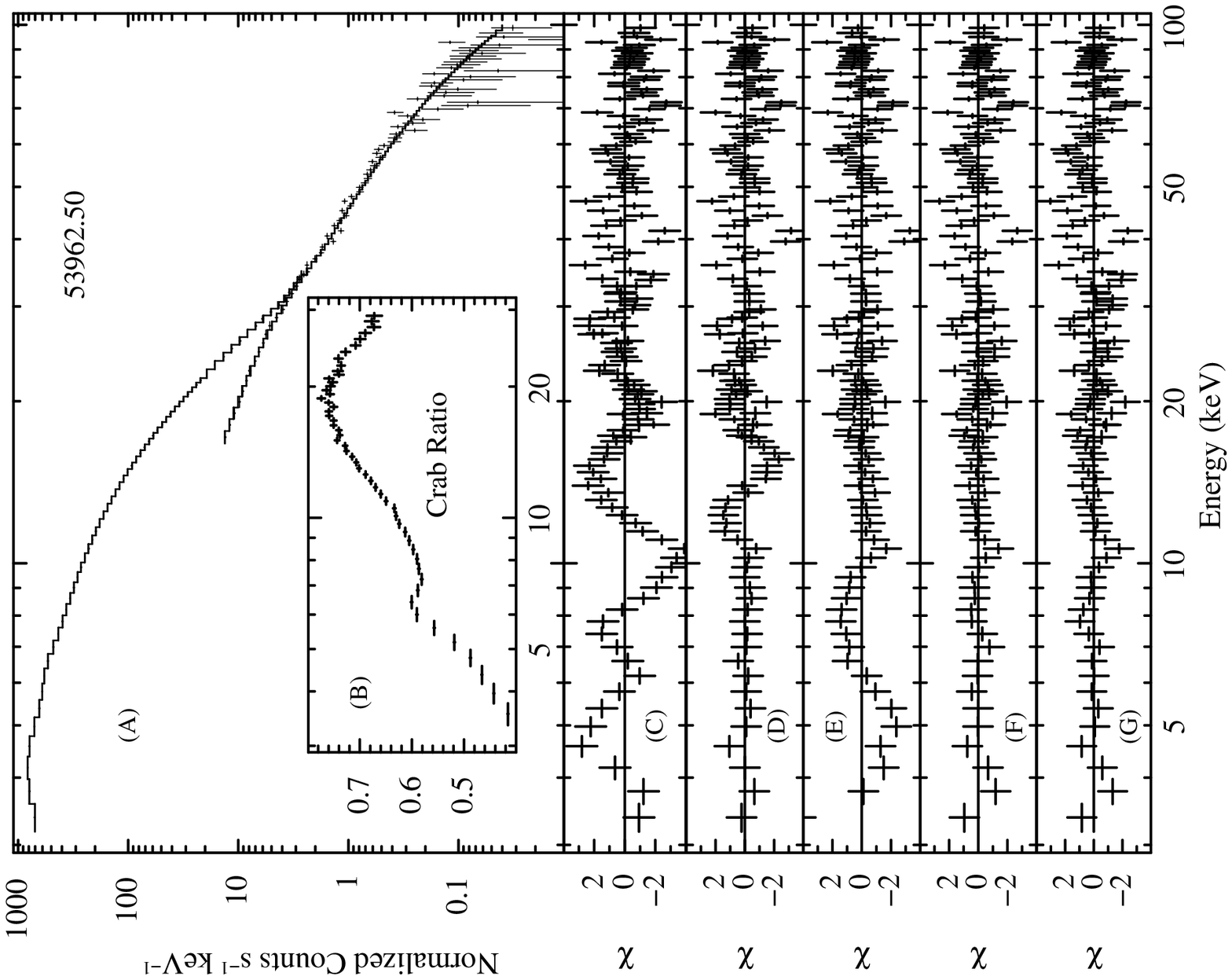}
\caption{The 3-100 keV energy spectra of EXO~2030+375 obtained from PCA and HEXTE detectors of {\it RXTE}
 at the peak of 2006 giant outburst (MJD 53962.50). Source spectra along with the best fitting model 
e.g. a partial covering high energy cutoff model along with a Gaussian for iron emission line (panel~A) 
and corresponding spectral residual (panel~G) are shown. Panels~C, D, E and F indicate the spectral 
residuals obtained by fitting pulsar spectra with a (i) high energy cutoff model, (ii) high energy 
cutoff model with a blackbody, (iii) CompTT with a blackbody, and (iv) a partial covering CompTT model
 with blackbody component, respectively, along with interstellar absorption component and a Gaussian 
function for iron emission line at 6.4 keV. Any signature of cyclotron absorption line at previously
 reported value of $\sim$11 keV is not seen in the spectral residuals. The Crab-ratio (panel~B) also
 did not show any such feature in 10-20 keV range.} 
\label{fig5}
\end{figure}

It has been recently found that the continuum of Be/X-ray binary pulsars is noticeably affected by
 the presence of additional matter at certain pulse phases during outbursts that can't be simply
 explained by a single absorber \citep{Naik2013, Jaisawal2016}. A partial covering absorption
 component along with the continuum model is generally used to describe the pulsar spectrum in
 phase-averaged as well as phase-resolved spectroscopy. In our fitting, a high energy cutoff
 power-law model was unable to fit the observed spectrum, specifically during bright phases of
 outbursts. Addition of a partial covering component improved the fitting further and yielded an 
acceptable value of reduced-$\chi^2$ ($\sim$1). The iron fluorescence emission line at $\sim$6.4~keV
 was also detected in the pulsar spectrum. A partially covering absorbed power law with high-energy
 cutoff model is mathematically expressed as 
%
\begin{eqnarray} \label{eq1}
\nonumber 
N(E) & = & {e^{-N{_{\mathrm H1}\sigma(E)}}}(K_{1}+K_{2}e^{-N_{\mathrm H2}\sigma(E)})
~{f(E)}
\end{eqnarray}

where
\begin{eqnarray}
\nonumber  f(E) & = & {E^{-\Gamma}}  \hspace{0.83in} for ~E < E_\mathrm c \\
\nonumber       & = & {E^{-\Gamma}}e^{- \left({E-E_\mathrm c}\over{E_\mathrm f}\right)} \hspace{0.39in}  for ~E > E_\mathrm c
\end{eqnarray}
%
where, {\it f(E)} represents the high energy cutoff power-law model with $\Gamma$ as the power-law
 photon index, $E_{c}$ and $E_{f}$ are the cutoff and folding energies in keV, respectively. 
The normalization constants $K_{1}$ \& $K_{2}$ are in the units of photon~keV$^{-1}$~cm$^{-2}$~s$^{-1}$.
 $N_{H1}$ \& $N_{H2}$ are the equivalent galactic hydrogen column density and the additional column
 density (in units of 10$^{22}$ atoms $cm^{-2}$), respectively. $\sigma(E)$ is the photoelectric 
absorption cross-section. The energy spectra of EXO~2030+375 along with the best-fitting model 
(top panel) and residuals (bottom panel) for six epochs of {\it RXTE} observations are shown in 
Figure~\ref{fig4}. Observation IDs and 3-30 keV pulsar luminosity during these epoch of 
observations are quoted in the figure. The best-fitted parameters obtained from spectral 
fitting of data obtained at these epochs are given in Table~\ref{table2}.

\begin{table*}
\centering
\caption{Best-fitting spectral parameters with 1$\sigma$ errors obtained from {\it RXTE/PCA} 
observations of EXO~2030+375 at six different luminosities. The best-fit model consists of 
a partial covering high-energy cutoff power-law model with a Gaussian component.}
\begin{tabular}{ |l | llllll}
\hline \\
Parameters                      &  \multicolumn{6}{c}{Observation IDs}    \\ 
\\
                                &70074-01-28-00     &80071-01-01-06     &91089-01-02-11     &92067-01-03-11	&91089-01-17-06		&91089-01-12-04  \\
\hline
N$_{H1}$$^a$                    &5.9$\pm$0.8         &6.2$\pm$0.5        &5.2$\pm$0.3	    &2.8$\pm$0.5        &1.5$\pm$0.4		&3.41$\pm$0.25 \\
N$_{H2}$$^b$                    &--         	     &--                 &--       	    &139.3$\pm$17.5     &166.5$\pm$5.5		&241.9$\pm$4.0 \\
Covering fraction               &--	             &--                 &--     	    &0.25$\pm$0.02	&0.35$\pm$0.01		&0.45$\pm$0.01 \\
Photon Index ($\Gamma$)         &1.72$\pm$0.06       &1.45$\pm$0.06      &1.34$\pm$0.03	    &1.43$\pm$0.05      &1.61$\pm$0.02		&1.80$\pm$0.03 \\
E$_{cut}$ (keV)	                &8.2$\pm$0.8         &7.6$\pm$0.4        &7.7$\pm$0.2       &8.0$\pm$0.3        &8.2$\pm$0.2		&7.8$\pm$0.2  \\
E$_{fold}$ (keV)		&31.9$\pm$5.6        &23.3$\pm$1.8	 &23.3$\pm$0.8      &21.2$\pm$0.9	&23.5$\pm$0.5		&25.7$\pm$0.8 \\
\\
{\it Emission lines } 
\\
Fe$~K\alpha$ line energy (keV)  &6.32$\pm$0.16       &6.50$\pm$0.11      &6.33$\pm$0.05     &6.51$\pm$0.08 	&6.43$\pm$0.03		&6.52$\pm$0.04 \\
Width of Fe line (keV)          &0.1	             &0.1                &0.1               &0.15               &0.31$\pm$0.07  	&0.1  \\
\\
Source flux  \\
Flux$^c$ (3-10 keV)   		&0.17$\pm$0.03      &0.88$\pm$0.09       &1.8$\pm$0.9       &6.6$\pm$0.7        &13.8$\pm$0.7            &25.0$\pm$1.5 \\
Flux$^c$ (10-30 keV)  		&0.16$\pm$0.02      &0.94$\pm$0.10       &2.2$\pm$0.1       &7.1$\pm$0.8        &12.7$\pm$0.6            &19.4$\pm$1.2  \\ 
Flux$^c$ (3-30 keV) 		&0.33$\pm$0.04      &1.82$\pm$0.19       &4.0$\pm$0.20      &13.7$\pm$1.4       &26.5$\pm$1.3            &44.4$\pm$2.6  \\
\\
Source Luminosity\\
L$_{X}$$^d$ (3-30 keV)              &0.20$\pm$0.03      &1.10$\pm$0.12       &2.38$\pm$0.14     &8.23$\pm$0.89       &15.96$\pm$0.90         &26.77$\pm$1.72 \\
\\
Reduced $\chi^2$(\it d.o.f)         &0.99 (24)          &1.01 (34)          &1.04 (46)        &1.00 (50)            &1.08 (53)               &1.11 (43) \\
\hline
\end{tabular}
\flushleft
Notes:
$^a$ : Equivalent hydrogen column density (in 10$^{22}$ atoms cm$^{-2}$ unit);
$^b$ : Additional hydrogen column density (in 10$^{22}$ atoms cm$^{-2}$ unit); 
$^c$ : Absorption corrected flux in unit of 10$^{-9}$  ergs cm$^{-2}$ s$^{-1}$; 
$^d$ : The 3-30 keV X-ray luminosity in the units of 10$^{37}$ ergs s$^{-1}$ assuming a distance of 7.1~kpc to the source.\\ 
\label{table2}
\end{table*}

\begin{figure*}
\centering
\includegraphics[height=7.0in, width=3.in, angle=-90]{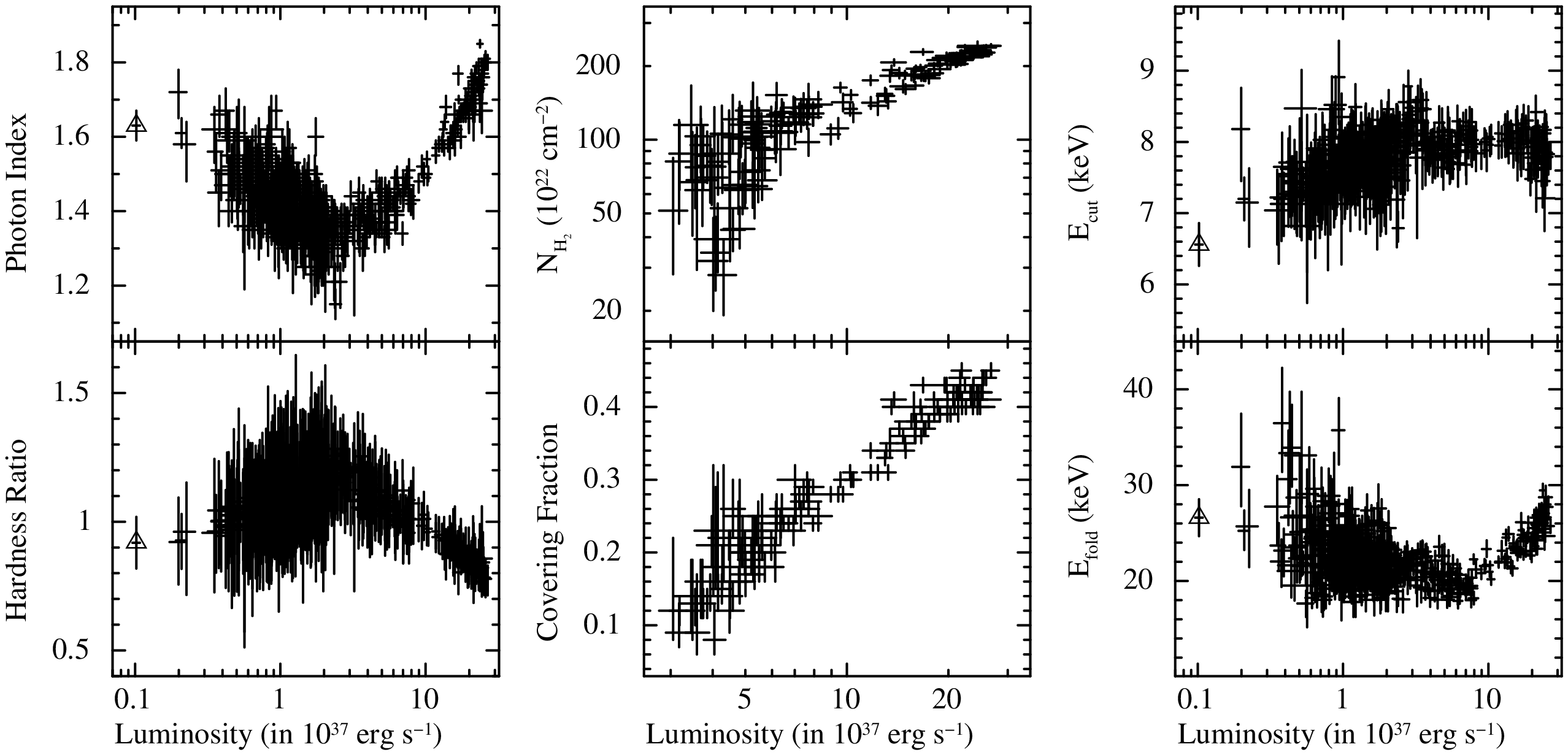}
\caption{Spectral parameters such as photon index (top left panel), additional column density (top middle panel),
 cutoff energy (top right panel), covering fraction (bottom middle panel) and folding energy (bottom right panel)
 obtained from the spectral fitting of {\it RXTE} observations of EXO~2030+375 with a partial covering high 
energy cutoff power-law model during Type~I and Type~II outbursts, are shown with the 3-30 keV luminosity.
 Hardness ratio (ratio between 10-30 keV flux and 3-10 keV flux) with luminosity is also shown in bottom
 left panel of the figure. The parameters from {\it NuSTAR} observation are marked with empty triangles. 
The error bars quoted for 1$\sigma$ uncertainties.} 
\label{fig6}
\end{figure*}

During 2006 June outburst, detection of a cyclotron absorption line like feature at $\sim$11 keV in
 the pulsar spectrum was reported \citep{Wilson2008}. We used same observation in the present study
 to investigate the cyclotron line feature in the pulsar spectrum. During this outburst, as the
 pulsar luminosity was very high, data from PCA and HEXTE detectors were used to get a broad-band 
spectral coverage. The 3-100 keV broad-band spectrum, obtained from the {\it RXTE} observation 
on MJD 53962.5, was fitted with a high energy cutoff power-law model yielding a poor fit with a
 reduced $\chi^2$ of $>$8. A broad absorption-like feature at $\sim$10~keV was detected in the 
spectral residual (panel~C of Figure~\ref{fig5}). Various combination of models such as high 
energy cutoff power-law, CompTT, NPEX along with other components such as blackbody or a partial
 absorber were used to test reliability of the reported line. Addition of a partial covering 
component to the above continuum models resolves the broad feature with a reduced-$\chi^2$ close to 1.
 Therefore, a high energy cutoff model along with a partial covering component was used as best-fit model
 in our analysis. We generated Crab-ratio to check the presence of absorption like feature in the pulsar spectrum.
 The Crab-ratio is obtained by normalizing the pulsar spectrum with the feature-less power-law spectrum of
 Crab pulsar to remove the presence of any uncertainties related to calibration and 
model (see also \citealt{Jaisawal2013}). This ratio showed a highly absorbed spectrum along with a 6.4 keV
 iron emission line below 10 keV. We did not find any signature of absorption feature in 10-20~keV range
 in the Crab-ratio (panel~B of Figure~\ref{fig5}). Spectral residuals obtained from fitting the pulsar
 spectral with different continuum models are shown in panels C, D, E and F of Figure~\ref{fig5}. 
The absence of any absorption like feature in 10-20 keV range can be clearly seen in above panels. 
To check the presence/absence of this absorption like feature, we fitted 3-79 keV broad-band 
spectrum of EXO~2030+375, obtained from a {\it NuSTAR} observation during an extended period of
 low activity in 2015 with a high energy cutoff model. Any emission/absorption like features 
were not seen in the spectrum. 

Spectral parameters such as power-law photon index, cutoff energy, folding energy, additional 
column density (NH$_{2}$), covering fraction, hardness ratio (ratio between 10-30 keV flux and 
3-10 keV flux) obtained from spectral fitting of all {\it RXTE} observations of EXO~2030+375 
are shown with corresponding 3-30 keV luminosity in Figure~\ref{fig6}. All these parameters 
showed intriguing trends with luminosity which had not been explored earlier. In the figure, 
one can notice that the values of power-law photon index are distributed in three distinct 
regions such as negative, constant, and positive correlations with source luminosity which 
suggest a direct measure of spectral transition in EXO~2030+375. At lower luminosity 
($\le$10$^{37}$erg~s$^{-1}$), the pulsar spectrum was relatively soft. A negative correlation 
between the power-law photon index and luminosity can be clearly seen for this regime. 
The value of photon index was found to varying between 1.2 and 1.8 (first panel of Figure~\ref{fig6}).
 When the luminosity was in the range of 2--4 $\times$10$^{37}$erg~s$^{-1}$, the distribution of 
values of photon index did not show any dependence on source luminosity. With increase in source luminosity,
 the photon index showed a positive correlation. It is, therefore, clear that the pulsar changes 
its spectral behavior with change in luminosity. Flux ratio (ratio between flux in 10-30 keV range and
 3-10 keV range) also showed smooth transition with increase in pulsar luminosity (left bottom panel of
 Figure~\ref{fig6}). As mentioned earlier, while fitting spectra from high flux level of the outbursts,
 a partially absorbed component was included in the fitting model. This was required in spectral fitting
 when the pulsar luminosity was above 3$\times$10$^{37}$erg~s$^{-1}$. In our spectral fitting, the maximum
 value of additional column density (NH$_2$) obtained was as high as 250$\times$10$^{22}$~cm$^{-2}$ which
 is significantly larger than the value of interstellar absorption column density in the source direction.
 From our fitting, the values of additional column density and covering fraction were found to be strongly
 luminosity dependent (middle panels of Figure~\ref{fig6}). The cutoff energy and folding energy did not
 show any noticeable changes with the pulsar luminosity (last panels of Figure~\ref{fig6}). 
The parameters obtained from {\it NuSTAR} observation are also included in the figure.


\subsection{A physical model to describe the pulsar continuum spectrum} \label{bwmodel}

To explore the physical properties of accretion column, we have fitted the 
Becker \& Wolff (BW) model with the phase averaged spectra of the pulsar. This model 
is proposed by Becker \& Wolff (2007) to explain the emission from accretion powered 
X-ray pulsars by considering the effects of thermal and bulk Comptonizations in accretion 
column. It has been successful in explaining the broadband spectra of bright X-ray pulsars 
such as 4U~0115+63 \citep{Ferrigno2009}, 4U~1626-67 \citep{Ai2017}, Her~X-1 \citep{wolff2016}. 
According to this model, seed photons originated due to bremsstrahlung, blackbody and 
cyclotron emissions undergo thermal and bulk Comptonization in the accretion column. 
Comptonization of these seed photons with highly energetic electrons lead to the power-law 
like resultant spectra with high energy exponential cutoff.

\begin{table*}
\centering 
\caption{Best-fitting spectral parameters with 1$\sigma$ errors obtained from {\it RXTE/PCA} 
and {\it RXTE/HEXTE} observations of EXO~2030+375 with BW model.}
\begin{tabular}{|l|cccccccr}
\hline 
& & \multicolumn{6}{c}{BW model Parameters} \\
& & \\\cline{3-8} \\
Obs-Ids  &Luminosity$^a$ &$\dot{M}$ &$\xi$ &$\delta$ &B &T$_{e}$ &r$_{0}$ &Reduced $\chi^{2}$ \\
  &(10$^{37}$ergs~s$^{-1}$) &(10$^{17}$~g~s$^{-1}$) &  & &(10$^{12}$)~G &(keV) & (m) & ({\it dof})\\
\hline \\
91089-01-12-04$^{\ddag}$ &\(28.04 \pm 0.84\) &14.96 &\(1.88 \pm 0.03\) &\(14.38 \pm 1.01\) &\(4.25 \pm 0.20\) &\(3.95 \pm 0.11\) &177.0 &1.22(103) \\
91089-01-12-05 &\(28.22 \pm 0.38\) &14.82 &\(1.77 \pm 0.02\) &\(17.57 \pm 1.20\) &\(4.23 \) &\(3.42 \pm 0.10\) &165.0 &1.04(80)\\
91089-01-15-00 &\(27.12 \pm 0.53\) &14.59 &\(1.82 \pm 0.02\) &\(16.82 \pm 1.19\) &\(4.58 \) &\(3.83 \pm 0.10\) &158.0 &1.05(104) \\
91089-01-14-03 &\(26.36 \pm 0.49\) &13.88 &\(1.86 \pm 0.03\) &\(17.03 \pm 2.50\) &\(4.34 \pm 0.25\) &\(4.01 \pm 0.16\) &141.0 &1.02(85) \\
91089-01-09-03 &\(19.39 \pm 0.39\) &10.44 &\(1.98 \pm 0.03\) &\(11.24 \pm 0.91\) &\(4.66 \pm 0.22\) &\(4.53 \pm 0.12\) &138.0 &1.09(105)\\
91089-01-09-00 &\(17.11 \pm 0.65\) &9.21 &\(2.06 \pm 0.05\) &\(9.64 \pm 2.75\) &\(5.29 \) &\(5.15 \pm 0.40\) &116.0 &1.09(104)\\
91089-01-08-01 &\(13.40 \pm 0.48\) &7.22 &\(2.13 \pm 0.05\) &\(10.48 \pm 1.10\) &\(5.59\) &\(5.61 \pm 0.14\) &105.0 &1.08(107) \\
91089-01-07-00 &\(11.63 \pm 0.58\) &6.26 &\(2.43 \pm 0.17\) &\(4.22 \pm 0.72\) &\(4.85 \pm 0.35\) &\(5.56 \pm 0.31\) &98.5 &1.13(98) \\
92067-01-03-13 &\(10.09 \pm 0.54\) &5.43 &\(2.44 \pm 0.16\) &\(4.17 \pm 0.54\) &\(5.18 \pm 0.36\) &\(5.97 \pm 0.40\) &74.0 &1.09(90) \\
92067-01-03-11 &\(8.17 \pm 0.98\) &4.40 &\(2.36 \pm 0.29\) &\(4.30 \pm 1.16\) &\(5.33 \pm 0.35\) &\(6.14 \pm 0.43\) &35.6 &1.16(95) \\
92067-01-03-00 &\(6.86 \pm 1.47\) &3.69 &\(3.11 \pm 0.73\) &\(2.09 \pm 1.07\) &\(5.03 \) &\(6.32 \pm 0.16\) &28.2 &1.10(99) \\
93098-01-03-05 &\(5.72 \pm 0.70\) &3.25 &\(1.91 \pm 0.16\) &\(7.90 \pm 1.38\) &\(4.85 \) &\(4.85 \pm 0.41\) &26.0 &1.08(87)\\
93098-01-01-01 &\(4.86 \pm 0.46\) &2.79 &\(2.08 \pm 0.15\) &\(5.50 \pm 1.10\) &\(4.35 \pm 0.21\) &\(4.88 \pm 0.27\) &25.5 &1.05(111) \\
92422-01-25-06$^{\ddag}$ &\(4.74 \pm 0.74\) &2.66 &\(3.01 \pm 1.19\) &\(2.43 \pm 1.20\) &\(5.39 \pm 0.45\) &\(6.68 \pm 0.36\) &24.5 &1.11(101)\\
93098-01-01-00 &\(3.90 \pm 0.83\) &2.34 &\(5.08 \pm 1.24\) &\(0.97 \pm 0.35\) &\(4.78 \) &\(6.32 \pm 0.07\) &19.6 &1.08(102)\\
92422-01-05-04 &\(3.82 \pm 0.84\) &2.22 &\(7.37 \pm 3.18\) &\(0.64 \pm 0.21\) &\(4.82 \) &\(6.53 \pm 0.05\) &19.2 &0.99(96)\\
93098-01-02-04 &\(3.71 \pm 1.08\) &1.98 &\(7.26 \pm 2.86\) &\(0.62 \pm 0.27\) &\(4.70 \) &\(6.34 \pm 0.07\) &23.4 &1.03(89) \\
92422-01-14-03 &\(3.07 \pm 0.75\) &1.74 &\(6.11 \pm 1.95\) &\(0.77 \pm 0.29\) &\(4.50 \) &\(6.09 \pm 0.07\) &17.5 &1.07(90)\\
80071-01-01-11 &\(1.41 \pm 0.01\) &0.81 &\(6.49 \pm 1.05\) &\(0.66 \pm 0.13\) &\(3.74 \) &\(5.24 \pm 0.02\) &17.5 &1.06(87)\\
80071-01-01-06 &\(1.20 \pm 0.01\) &0.67 &\(8.15 \pm 2.50\) &\(0.52 \pm 0.20\) &\(3.77 \) &\(5.29 \pm 0.03\) &16.0 &1.08(86)\\
80071-01-01-00 &\(1.14 \pm 0.01\) &0.64 &\(7.11 \pm 1.65\) &\(0.60 \pm 0.17\) &\(4.02 \) &\(5.69 \pm 0.03\) &18.4 &1.19(97)\\
80071-01-01-01$^{\ddag}$ &\(0.88 \pm 0.01\) &0.48 &\(7.89 \pm 1.80\) &\(0.57 \pm 0.15\) &\(3.75 \) &\(5.32 \pm 0.02\) &17.4 &1.12(87)\\
80071-01-01-020 &\(0.92 \pm 0.01\) &0.44 &\(10.29 \pm 2.58\) &\(0.39 \pm 0.16\) &\(3.69 \) &\(5.28 \pm 0.01\) &15.0 &1.03(72) \\
\hline
\end{tabular}
\flushleft
Notes:
$^a$ : The 3-70 keV luminosity in the units of 10$^{37}$ ergs s$^{-1}$ by assuming a distance of 7.1~kpc. \\
$^\ddag$: indicates the observation ids from which extracted spectra are shown in Figure~\ref{fig7}. \\
\label{table3}
\end{table*}

\begin{figure*}
 \begin{center}$
 \begin{array}{cccc}
 \includegraphics[height=5.5 cm,angle=-90]{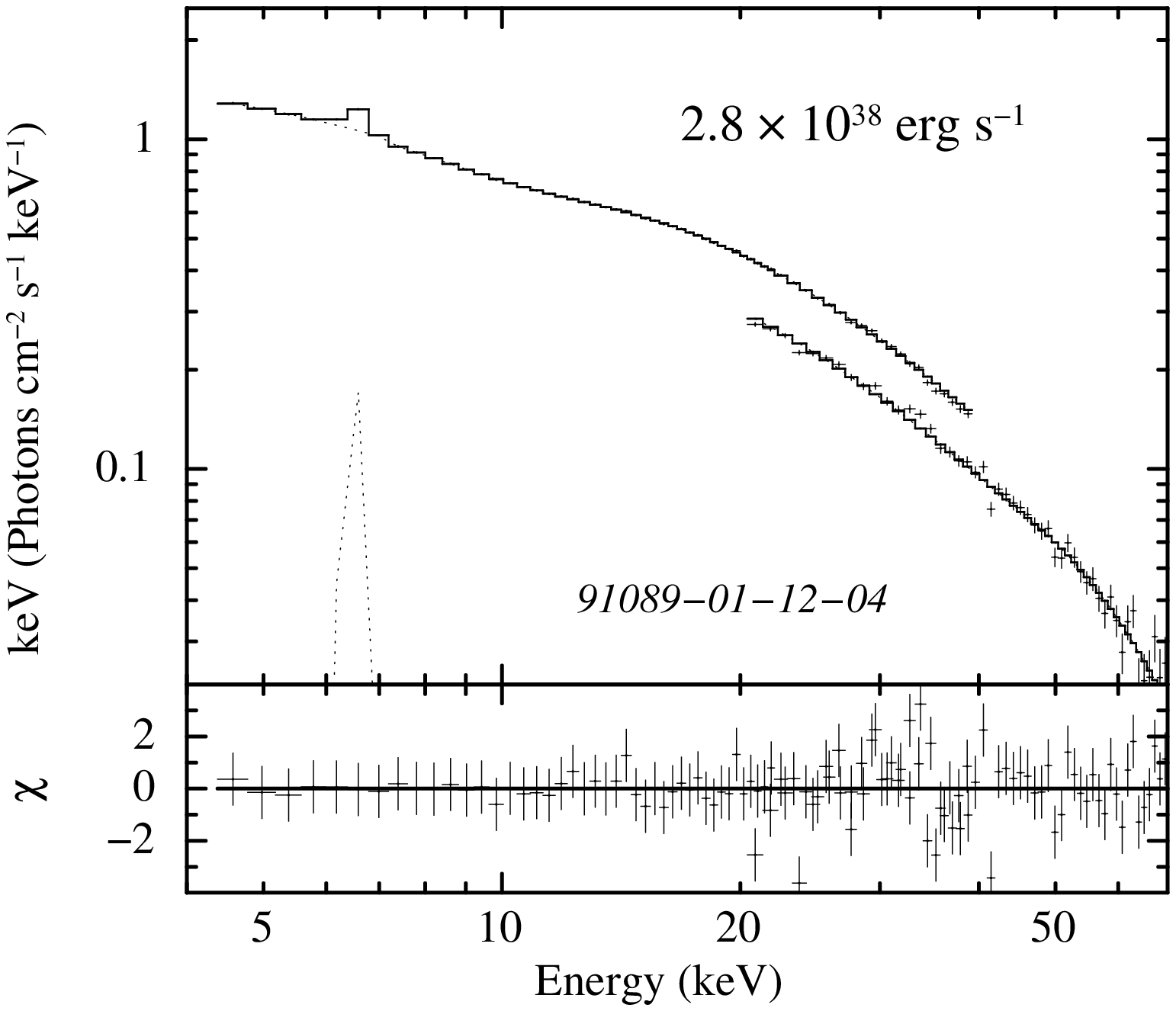} & 
 \includegraphics[height=5.5 cm,angle=-90]{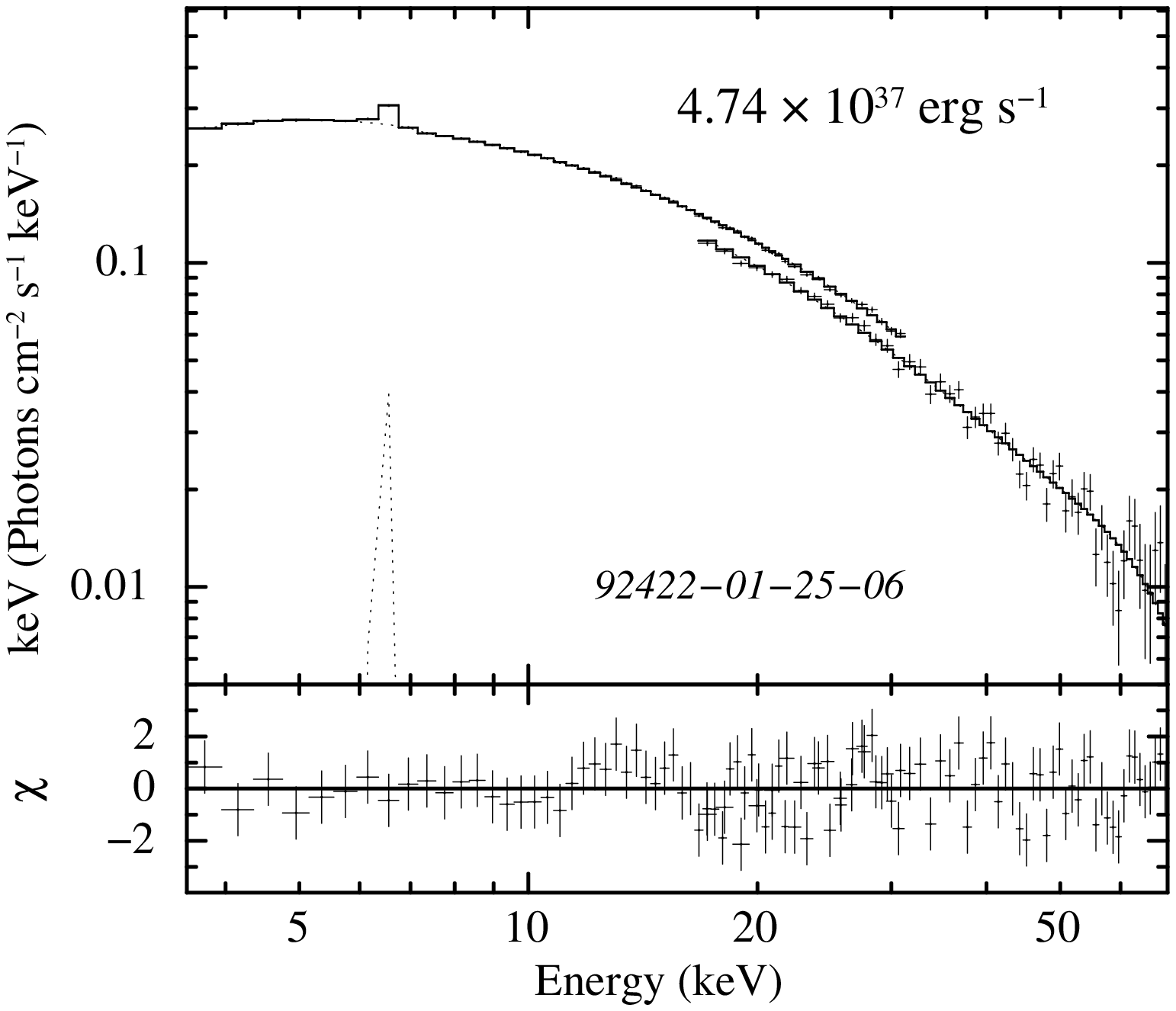} & 
 \includegraphics[height=5.5 cm,angle=-90]{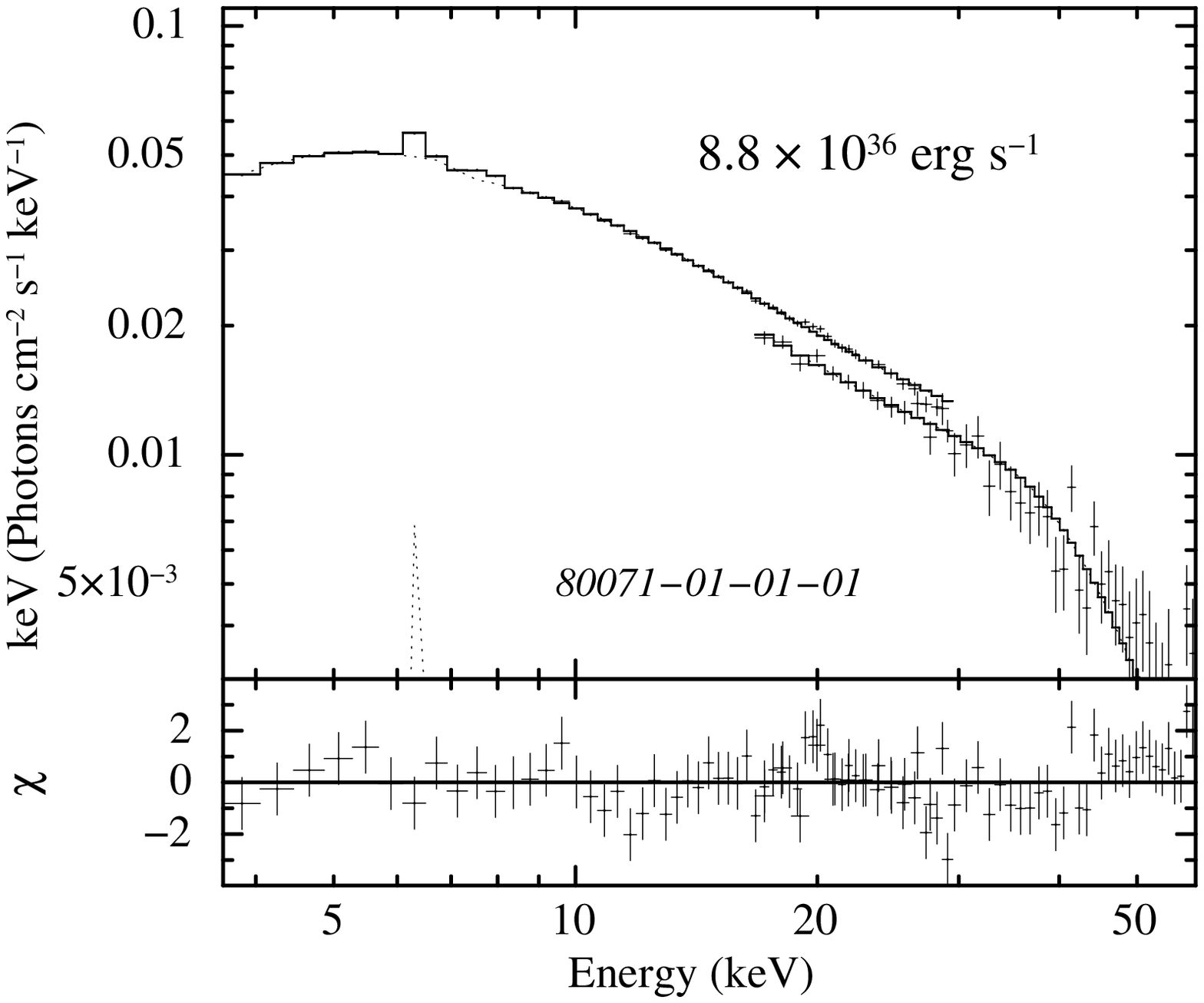} & \\
 \end{array}$
 \end{center}
\caption { Phase-averaged energy spectra of EXO~2030+375 at three distinct 
luminosities obtained during Type-I and Type-II X-ray outbursts obtained from PCA 
and HEXTE detectors. The spectra of pulsar was fitted with BW model \citep{Ferrigno2009} 
along with an iron line at $\sim$6.4 keV and partial covering component (top panel of 
figure). Corresponding spectral residuals are shown in bottom panels of the figure. Any 
absorption like feature was not seen in 10--20 keV energy range of the pulsar spectra.}
 \label{fig7}
 \end{figure*}

Using this model, we have described the 3-70~keV broadband phase-averaged spectra of EXO~2030+375 
at 23 different luminosity epochs, covering the range of 10$^{36}$--10$^{38}$~erg~s$^{-1}$.
For a canonical neutron star mass and radius, the BW model has six free parameters, i.e.
the diffusion parameter $\xi$, the ratio of bulk to thermal Comptonization $\delta$,
the column radius $r_{0}$, mass accretion rate $\dot{M}$, electron temperature $T_{e}$
and the magnetic field strength $B$. Among these parameters, the mass accretion rate $\dot{M}$ 
was estimated by using the observed source flux obtained from high energy cutoff empirical 
model and considering a source distance of 7.1~kpc \citep{Wilson2002}. Since the column radius 
strongly depends on accretion rate (see eq. 112 of \citealt{Becker2007}), the parameter 
$\dot{M}$ was fixed at a given value while fitting the spectra. After fitting, the column 
radius was also fixed for getting better constraints on other spectral parameters.
This was done carefully by analyzing the 2-D contour plots between $r_{0}$ \& $\dot{M}$, 
as similar to \citet{Ferrigno2009}. The other BW model components such as normalizations 
of Bremsstrahlung, cyclotron and blackbody seed photons were also kept fixed as suggested 
in the BW\_cookbook\footnote{\url {http://www.isdc.unige.ch/~ferrigno/images/Documents/BW\_distribution/BW\_cookbook.html}}.
A partial covering component as required in empirical models was also needed to explain 
the absorbed spectra of the pulsars during bright outbursts. The values of additional 
column density and covering fraction obtained from BW model were found to be consistent 
with the values obtained with the high energy cutoff power-law model. Therefore, we have 
not discussed these parameters in this section. An iron fluorescence line at $\sim6.4$~keV 
was also added in the continuum. Spectral parameters obtained after best fitting the pulsar 
spectra with BW model are given in Table~\ref{table3}. The values of reduced $\chi^2$ 
obtained from our fitting, as given in Table~\ref{table3}, showed that the BW continuum model
fits the data well in a wide luminosity range. For three different values of pulsar luminosity, 
broadband energy spectra of the pulsar from PCA and HEXTE detectors, along with 
best-fitted BW continuum model and Gaussian function for iron emission line are presented 
in top panels of the Figure~\ref{fig7}. The bottom panels in this figure show corresponding 
spectral residuals of best fitted model. It can be seen that the residuals obtained from fitting
the pulsar spectra with BW model did not show any evidence of presence of absorption like
feature. This finding also supports the non-detection of cyclotron line in EXO~2030+375, 
as discussed in the above section of the paper.

\begin{figure}
 \includegraphics[height=3.3in,width=5.2in,angle=-90]{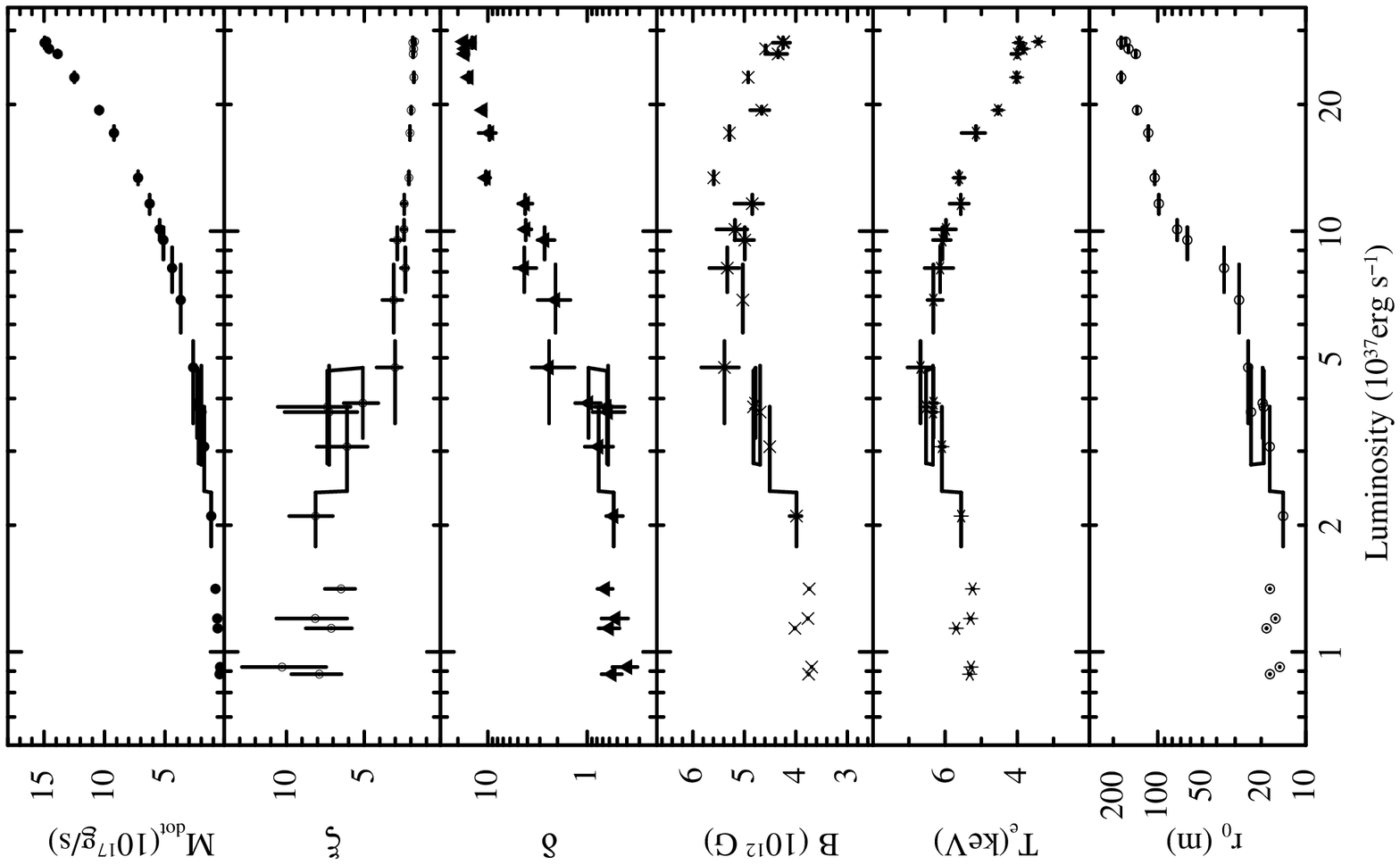} \\
 \caption{Spectral parameters obtained from the fitting of phase averaged 
 spectra of EXO~2030+375 with BW model at different luminosities. The top, second 
 and third panels of figure show the mass accretion rate, diffuse rate, and 
 the ratio of bulk to thermal Comptonization in accretion column, respectively.   
While the fourth, fifth and sixth panels indicate the luminosity variation of magnetic field, plasma 
temperature and column radius, respectively.}
 \label{fig8}
 \end{figure}

Luminosity dependent variations in the parameters obtained after fitting the pulsar spectra
with BW model are shown in Figure~\ref{fig8}. An interesting trend of parameter $\delta$  
with luminosity was noticed in the third panel of the figure. This parameter signifies 
the ratio of bulk to thermal Comptonization occurring in the accretion column. The value 
of $\delta$ was found close to unity at luminosity $\le$3--4$\times$10$^{37}$~ergs~s$^{-1}$. 
This indicates that the effects of thermal and bulk Comptonization are nearly same in accretion 
column  at lower luminosity of the pulsar. However, as the luminosity increases, bulk 
Comptonization starts playing a major role to column emissions and dominates over by a 
factor of 20 times as observed at lower luminosity. The column radii $r_{0}$ was found to 
strongly dependent on luminosity or mass accretion rate. Moreover, the diffusion parameter 
$\xi$ was also observed to vary with source intensity, showing a minimum value at higher 
luminosity. In addition to these, the electron plasma temperature was changing in the range
of 3 to 7 keV. The temperature showed a gradual increase up to luminosity 
4$\times$10$^{37}$~ergs~s$^{-1}$. Beyond this, a cooling of plasma temperature 
was observed. This may occur in the presence of strong radiation dominated accretion 
shock at which the infalling matter mostly bulk Comptonize the seed photons that 
carries plasma energy by diffusing through the side walls of accretion column. 
It leads to the settling of plasma in accretion column at lower temperature.  
This model also provides an opportunity to constrain the magnetic field 
of the pulsar in the range of $\sim$4--6$\times$10$^{12}$~G (see Figure~\ref{fig8}).
In some cases, magnetic field estimated from this model was found insensitive to
upper value, though their lower estimate was easily constrained in all these observations. 
Therefore, only best fitted values without error bars are quoted in Table~\ref{table3}.

\section {Discussion} 

Be/X-ray binaries show two types of X-ray outbursts such as Type~I and Type~II outbursts.
These outbursts are known to be due to capture of huge amount of matter by the neutron star from the 
circumstellar disk of the companion Be star at periastron passage (Type~I) or due to evacuation/truncation
 of Be circumstellar disk (Type~II). During these episodes, X-ray emission from the neutron star gets
 transiently enhanced by a factor of ten or more. Normal (Type~I) outbursts are found to be less luminous
 (10$^{36-37}$~erg~s$^{-1}$) whereas during giant (Type~II) outbursts, the luminosity reaches close
 to or above the Eddington luminosity of a neutron star \citep{Negueruela1998}. EXO~2030+375 is a 
unique Be/X-ray binary system that displays Type~I outbursts almost at each periastron passage. 
However, the pulsar showed Type~II outbursts only twice (in 1985 and 2006) since its discovery. 
Type~I outbursts are short lived, covering about 20-30\% of orbital period in contrast to giant
 outbursts which lasted for more than 3 months in both the occasions. The peak luminosity during 
normal outburst varies depending on the evolutionary state of the Be circumstellar disk. It is
 believed that the neutron star accretes matter from the circumstellar disk at the periastron
 passage and gives rise to Type~I X-ray outbursts. As the Type~II X-ray outbursts are very rare,
 the origin for Type~II outburst remains unclear. It is thought that dramatic expansion of the
 circumstellar disk around the Be star or instabilities in the circumstellar disk leads to such
 major events \citep{Okazaki2001, Reig2011}. Several analytical and numerical studies have been
 performed to understand the behavior of transient accretion by considering a misalignment between
 the orbital plane and the eccentric warped Be circumstellar disk \citep{Okazaki2013, Martin2014}.
 These studies showed that accretion time scales longer than the orbital period and higher luminosity
 as observed during Type~II outbursts are possible in such scenario. Using large number of {\it RXTE}
 observations of EXO~2030+375 during Type~I and Type~II outbursts, we have explored the evolution of
 pulse profile and its characteristics as a function of luminosity as well as type and phase of outbursts.
 The changes in spectral properties and its transition are also studied in detail in the paper.

\subsection{Pulse profiles}

Pulse profiles of Be/X-ray binary pulsars are generally complex due to the presence of multiple 
absorption dips at various spin phases of the pulsar. We have explored the evolution of pulse profiles of
 EXO~2030+375 over a wide range of luminosity, starting from $\sim$10$^{36}$ 
to 10$^{38}$ erg s$^{-1}$ by using large number of {\it RXTE} observations of the pulsar. The profiles 
are found to be strongly luminosity dependent which has also been reported from previous studies 
\citep{Parmar1989b, Klochkov2008, Naik2013, Naik2015}. One of the interesting results we obtained
 is that the profiles are similar in shape at identical luminosities irrespective of types and phases
 of outbursts. This indicates that the beam pattern of the pulsar does not depend on outburst characteristics,
 although distinct mechanism governs the origin of Type~I and Type~II outbursts. Accretion rate or 
luminosity is the vital factor that decide the shape of pulse profile of the pulsar. At lower luminosity
 ($\le$10$^{37}$ erg s$^{-1}$; sub-critical regime), the X-ray emission is believed to be originated 
from the hot spot along the accretion column in the form of pencil beam (\citealt{Sasaki2010} and
 references therein). This beam geometry leads to the formation of a single peaked profile as seen
 in our study at lower luminosity. In presence of a radiation dominated shock at higher luminosity,
 the profile changes from single to double peaked. A mixture of 
pencil and fan beam patterns can contribute to the double peak structure of pulse profiles which
 is clearly seen in EXO~2030+375 in the luminosity range of $\sim$(3-12)$\times$10$^{37}$~erg~s$^{-1}$. 
During the giant outbursts, mass accretion rate increases beyond the critical luminosity that shifts 
the height of raditaive shock in the accretion column. As a result, accreted matter is obstructed by 
dominating radiation pressure above the shock \citep{Becker2007}. The photons beyond this point mostly
 diffuses through the side wall of the accretion column, forming a fan-beam geometry \citep{Klochkov2008}.
 Therefore, at higher luminosity ($>$1.2$\times10^{38}$~erg~s$^{-1}$), the pulse profiles of the pulsar
 could be purely due to the fan beam pattern.   

The energy dependent pulse profiles in EXO~2030+375 has also been reported earlier 
\citep{Reig1999, Klochkov2008, Naik2013, Naik2015}. The shape of the pulse profiles of the pulsar
 is complex due to the presence of multiple peaks/dips in soft X-rays. A single peak profile is 
generally seen at hard X-rays. \citet{Sasaki2010} modeled the energy resolved pulse profiles during
 a giant outburst to understand the emission components from the magnetic poles of the pulsar.
 An asymmetric profile was explained by considering a moderate distortion in the magnetic field,
 which can locate one of the accretion column at relatively close to the line of sight. The composite 
emissions from both the poles resulted an asymmetry in the profile due to deformity in the location of columns.
 {\it Suzaku} observations of the pulsar during 2007 and 2012 Type~I outbursts showed nearly symmetrical profiles
 along with multiple absorption dips at certain pulse phases \citep{Naik2013, Naik2015}. These dips showed strong
 dependence on energy and luminosity. Presence of additional matter in the form of narrow streams that are phase
 locked to the pulsar is believed to be the cause of these features in pulse profiles. Such absorption like 
features are also seen in pulse profiles of other Be/X-ray binary pulsars such as A~0535+262 \citep{Naik2008},
 GRO~J1008-57 \citep{Naik2011}, and GX 304-1 \citep{Jaisawal2016} during outbursts. During 2007 outburst of
 EXO~2030+375, these dips were present in the profiles up to $\sim$70 keV \citep{Naik2013}.
 Phase resolved spectroscopy confirmed the presence of additional dense matter at same phases of the profile.
 In addition to absorption dips, sometimes a narrow and sharp dip-like feature was detected in the pulse
 profiles at low luminosity \citep{Ferrigno2016}. This feature was interpreted as due to self absorption from
 the accretion column. This peculiar feature was also detected in the pulse profiles of EXO~2030+375 and 
found to be luminosity dependent. The feature was present in the pulse profiles when the pulsar luminosity
 was below 4$\times10^{37}$~erg~s$^{-1}$ (critical luminosity regime). It is probable that at lower luminosity,
 the pencil beam propagating across the magnetic axis interacts with the accretion column directly and 
produces a dip-like structure in the pulse profile. A significant contribution from fan beam may change 
the emission geometry and lead to the absence of this feature beyond the critical luminosity, as seen in our study. 

\subsection{Spectroscopy} \label{spec-discuss}

The broadband energy spectrum of accretion powered X-ray pulsar is known to be originated due to the inverse Comptonization
 of soft X-ray photons emitted from the hot spots on the surface \citep{Becker2007}. The continuum is generally described 
with simple models such as high energy cutoff power law, NPEX, exponential cutoff power law etc. despite of the
 complex phenomenon occurring in the accretion column. A physical model, known as BW model, was also used 
in our study to understand the spectral properties of the pulsar at different luminosity levels. We carried out 
spectral analysis of a large number of {\it RXTE} pointed observations of Be/X-ray binary pulsar EXO~2030+375, 
spanned over a decade by using a high energy cutoff power-law model along with a partial absorbing component 
and a Gaussian function for the 6.4 keV iron emission line. The {\it RXTE} observations of the pulsar provided 
opportunity to trace spectral evolution of the pulsar at various luminosity levels during Type~I and Type~II 
X-ray outbursts since 1996 to 2011. Parameters obtained from the spectral fitting showed very interesting 
variation with luminosity. The photon index was found to exhibit three distinct patterns with luminosity 
indicating signatures of spectral transition between sub-critical and super critical states. All three 
regimes are reflected in the pattern of the pulse profiles and are interpreted as due to different beam 
pattern at three different luminosity ranges.

A negative correlation was seen between power-law photon index and pulsar luminosity in sub critical 
regime (below the critical luminosity) where spectrum was relatively hard. In this condition,
 the broadband X-ray emission is considered to be originated from a hot mount on the neutron star 
surface \citep{Basko1976, Becker2012}. Critical luminosity is associated with the transition between two 
accretion scenarios. In our study, we detected a plateau like region in the distribution of power-law photon
 index in luminosity range of $\sim$(2-4)$\times$10$^{37}$~erg s$^{-1}$. This luminosity range can be
 considered as the critical luminosity for EXO~2030+375. The critical luminosity regime has been explored
 for other accretion powered X-ray pulsars such as 4U~0115+63, V~0332+53, Her~X-1, A~0535+26 and GX~304-1
 in luminosity range of $\sim$(2-8)$\times$10$^{37}$~erg s$^{-1}$ (\citealt{Becker2012} and reference therein).
 A positive correlation between photon index and luminosity was detected above the critical luminosity.
 This occurs because of dominating role of shock in the accretion column which effectively reduces the
 velocity of energetic electrons. In this case,  pulsar spectrum appears soft due to lack of bulk Comptonization
 of photons with accreting electrons \citep{Becker2012}. A positive correlation between photon index and luminosity,
 therefore, is observed in super-critical regime. During 1985 giant outburst, the photon index was also proportional
 to luminosity, indicating that the pulsar was accreting above the critical limit \citep{Reynolds1993}.

The hardness ratio (ratio between 10-30 keV flux and 3-10 keV flux) also showed similar kind of transition with
 luminosity. This showed increasing trend till the pulsar luminosity reaches its critical value beyond which
 a decreasing trend was observed. It showed that the pulsar emission was  relatively hard in the sub-critical
 luminosity region. A softening in the spectrum was observed above the critical luminosity as discussed in
 above section. The folding energy was found to vary with luminosity. This spectral parameter represents
 the plasma temperature in the emission region. At lower luminosity, the value of folding energy was
 relatively constant although a few high values (with large errorbars) were also evident. The high
 values in sub-critical luminosity regime may correspond to the deep regions of accretion column.
 The value of folding energy was increasing beyond the critical luminosity. It is expected that
 increasing mass accretion in super-critical region can produce the high temperature plasma in the presence of shock.

The magnetic field of the pulsar can be investigated by using observed cyclotron resonance scattering features
 in the broad-band spectrum. Cyclotron resonance scattering features appear due to the resonant scattering of
 electrons with photons in the presence of magnetic field \citep{Caballero2012}. These absorption like features
 appear in the hard X-ray spectrum of accretion powered X-ray pulsars with magnetic field in the order of 
10$^{12}$ G. The detection of these features allow us to directly estimate the magnetic field of pulsar.
 Detection of a cyclotron absorption line at $\sim$11~keV in the pulsar spectrum obtained from {\it RXTE}
 observation was reported earlier \citep{Wilson2008}. Using same data set, we attempted to explore the
 cyclotron line feature further. However, our results showed that this feature was model dependent, only seen
 in a single cutoff based model. This discards the detection of cyclotron scattering feature in EXO~2030+375. 
The feature was also not detected in the pulsar spectra obtained from {\it NuSTAR} observations, 
though it was carried out during low intensity phase. Moreover, our studies based on BW model 
showed a constrain on pulsar magnetic field which is in the range of $\sim$4--6$\times$10$^{12}$~G. 
For such strength of magnetic field, a cyclotron feature is expected to observed in the 40--60 keV 
energy range of pulsar spectrum. We have not observed any such features in the above energy 
range using {\it RXTE} observations, although a claim of cyclotron line at $\sim$36 or 63~keV
have been made earlier in the pulsar \citep{Reig1999, Klochkov2008}. With high capability of new 
generation satellite such as {\it Astrosat} and {\it NuSTAR}, the cyclotron line feature can be 
investigated during an intense X-ray outburst.

\section {Summary and Conclusion}

We have carried out a detailed timing and spectral analysis of transient Be/X-ray binary pulsar 
EXO~2030+375 during several Type~I and Type~II outbursts during 1996-2011 with {\it RXTE}.
 An absorbed power-law modified with a high-energy cutoff along with a partial absorber and 
a Gaussian component was used to explain the 3-30 keV energy spectrum of the pulsar. The pulse
 profiles were found to be strongly luminosity dependent at wide range of $\sim10^{36 - 38}$~erg~s$^{-1}$.
 Observed changes in the shape of pulse profiles are attributed to the change in emission geometry of pulsar.
 Pulsar emission was dominated by pencil beam at lower luminosity which switched to fan beam at higher luminosity.
 The different shape of pulse profiles at both the extremes are interpreted as due to different beam patterns
. However, a mixture of both patterns are seen in the critical luminosity regime. The profiles were observed
 to be independent of Type~I and Type~II outbursts and their phases. Spectral parameters also showed the
 signatures of emission transition. Based on luminosity, a transition from sub-critical to super-critical
 regime is seen in the photon index. These changes are explained in terms of changes in the emission
 geometry across the critical luminosity. At the brighter phases the presence of additional matter nearby
 the pulsar is observed due to effect of higher accretion rate. Based on the physical modeling of the 
continuum spectrum, the magnetic field of pulsar can be estimated to be  $\sim$4--6$\times$10$^{12}$~G.

\section*{Acknowledgments}
We thank the referee for his/her constructive suggestions that improved the 
paper. The research work at Physical Research Laboratory is funded by the 
Department of Space, Government of India. This research has made use 
of data obtained through HEASARC Online Service, provided 
by the NASA/GSFC, in support of NASA High Energy Astrophysics Programs.


\label{lastpage}
\end{document}